\documentclass[twocolumn,showpacs,floatfix,pre,superscriptaddress]{revtex4-1}

\usepackage{graphicx} 
\usepackage{dcolumn} 
\usepackage{epsf} 
\usepackage{amsmath}
\usepackage{amssymb}
\usepackage{bm} 
\usepackage{setspace}
\usepackage{color}

\usepackage{cmap}
\usepackage{lmodern}
\usepackage[english]{babel}

\newcommand{\sscf}{{\it SSCF\;\,}}
\newcommand{\ppcf}{{\it PPCF\;\,}}
\newcommand{\sscfs}{{\it SSCFs\;\,}}
\newcommand{\ppcfs}{{\it PPCFs\;\,}}

\begin{document}


\title{Contribution to Viscosity from the Structural Relaxation via the Atomic Scale Green-Kubo Stress Correlation Function}
\author{V.A.~Levashov}
\affiliation{Technological Design Institute of Scientific Instrument Engineering,
 Novosibirsk, 630058, Russia}


\begin{abstract}
We studied the connection between the structural relaxation and viscosity for a binary model of repulsive particles in the supercooled liquid regime. 
The used approach is based on the decomposition of the macroscopic Green-Kubo stress correlation function into the correlation functions between the atomic level stresses. 
Previously we used the approach to study an iron-like single component system of particles. 
The role of vibrational motion has been addressed through the demonstration of the relationship between viscosity and the shear waves propagating over large distances. 
In our previous considerations, however, we did not discuss the role of the structural relaxation. 
Here we suggest that the contribution to viscosity from the structural relaxation can be taken into account through the consideration of the contribution from the atomic stress auto-correlation term only. 
This conclusion, however, does not mean that only the auto-correlation term represents the contribution to viscosity from the structural relaxation. 
Previously the role of the structural relaxation for viscosity has been addressed through the considerations of the transitions between inherent structures and within the mode-coupling theory by other authors. 
In the present work, we study the structural relaxation through the considerations of the parent liquid and the atomic level stress correlations in it.
The comparison with the results obtained on the inherent structures also is made.
Our current results suggest, as our previous observations, that in the supercooled liquid regime the vibrational contribution to viscosity extends over the times which are much larger than the Einstein's vibrational period and much larger than the times which it takes for the shear waves to propagate over the model systems. 
Besides addressing the atomic level shear stress correlations, we also studied correlations between the atomic level pressure elements.
\end{abstract}

\today

\maketitle


\section{Introduction}

In simple considerations of supercooled liquids it is quite common to draw a parallel 
with the Maxwell viscoelastic model \cite{MaxwellJC18671,BlandDR19601,HansenJP20061}. 
In this model a spring and a viscous dump element are connected in series. 
If a strain is applied suddenly to the system then it instantly stretches 
or compresses the spring, storing the potential energy in it. 
The energy stored in the spring results in its tension or compression. 
Later, the energy stored in the spring is dissipated in the viscous dump element exponentially in time. 
In drawing the parallel with supercooled liquids it is assumed that a suddenly strained 
supercooled liquid stores in itself the potential energy due to elastic deformation and 
that later this elastically stored potential energy slowly dissipates through some 
microscopic processes that occur in the supercooled liquid. 
In the framework of this parallel, it is natural to assume that the elastic energy stored 
in the supercooled liquid should be associated with its stressed state.

In several recent publications it has been discussed that the inherent structures of the model 
liquids have non-zero macroscopic stress states \cite{HarrowellP20121,IlgP20151,HarrowellP20161}. 
It has been demonstrated that these non-zero stress states are related to the boundary 
conditions which put constraints on the allowable configurational states of the systems \cite{IlgP20151}.
Given of the Maxwell model discussed above, it is natural to assume that the non-zero 
macroscopic stresses of the inherent states manifest about the potential energy elastically stored in the inherent states. 
Thus different inherent states should correspond to the different states 
of the elastic energy in nearly a tautological agreement with the definition of the inherent states. 
From this perspective transitions between the inherent states 
should be associated with dissipation or accumulation of the stored elastic energy. 

Over the several decades the mechanism of the stress relaxation in supercooled liquids has been extensively discussed 
\cite{HansenJP20061,HarrowellP20121,IlgP20151,HarrowellP20161,McQuarrie19761,Green19541,Kubo19571,Helfand19601,Hoheisel19881,
Woodcock19911,Woodcock20061,Steele19951,Picard20041,Leonforte20041,Tanguy20061,Tsamados20081,Leporini20121,Lemaitre20151}.
Many of these considerations address stress relaxation through considerations of the stress correlations relevant 
to the Green-Kubo expression for viscosity \cite{HansenJP20061,HarrowellP20121,IlgP20151,
HarrowellP20161,McQuarrie19761,Green19541,Kubo19571,Helfand19601,Hoheisel19881,
Woodcock19911,Woodcock20061,Steele19951,Leporini20121}.
In particular, it has been shown that in deeply supercooled liquids (parent liquids) relaxation of the Green-Kubo stress 
correlation function at large times corresponds to the relaxation of the Green-Kubo stress correlation function calculated 
on the inherent structures (i.e., on the structures obtained through ``an instant quench" from the parent liquid structures)
\cite{HarrowellP20121,IlgP20151,HarrowellP20161,Leporini20121}. 
Another direction in the considerations of the stress relaxation concerns the angular 
dependence of changes in the stress fields generated by local structural 
transformations \cite{Picard20041,Tanguy20061,Tsamados20081,Lemaitre20151,Chattorai20131,Puosi20161}.

In our previous considerations of the atomic stress correlations relevant to the Green-Kubo expression 
for viscosity we investigated the connection between viscosity and propagating shear 
stress waves \cite{Levashov20111,Levashov2013,Levashov20141,Levashov2014B}. 
In particular, we found that the contribution to viscosity associated with the shear 
stress waves is very non-local in the model liquid even at temperatures well above the melting temperature. 
In those previous considerations the role of the structural relaxation in liquids has not 
been discussed \cite{Levashov20111,Levashov2013,Levashov20141,Levashov2014B}. 
One (and, probably, the major) purpose of the current paper is to elucidate 
how the structural relaxation is expressed in the atomic stress correlation function. 
Previously the connection between the structural relaxation and viscosity 
has been addressed from the perspective of
the inherent structures \cite{Leporini20121,HarrowellP20121,HarrowellP20161}. 
Here we suggest that the structural relaxation in the Green-Kubo stress correlation 
function of a parent liquid at times larger 
than few Einstein's vibrational periods can essentially be taken into account through the 
consideration of the atomic stress autocorrelation term only.
Another purpose of the paper is to assess the generality of the previously 
obtained results through consideration of a different system. 
The binary model system that we consider in this paper can be supercooled to 
significantly lower temperatures than the single component system that 
has been studied in Ref.\cite{Levashov20111,Levashov2013,Levashov20141,Levashov2014B}.  
Finally, we studied the atomic scale pressure-pressure correlation function. 
Behavior of this correlation has not been studied previously from the perspective 
of the atomic scale correlations. 
We found that the pressure-pressure correlations (longitudinal waves) are 
very well pronounced in the atomic scale pressure-pressure correlation function.

Our current interpretations of the results are strongly influenced by Ref.\cite{Petravic20051}.
In Ref.\cite{Petravic20051} the vibrational contribution to the crystals' viscosity has been considered.
It has also been stated there that situation in supercooled liquids is different in a 
sense that there is an additional contribution to the liquids viscosity from the structural relaxation.
While these results at present appear to be quite natural, we mention them because 
in Ref.\cite{Petravic20051} the vibrational contribution to viscosity has been explicitly and carefully discussed. 
Those discussions appear to be well aligned with our previous and the current results. 
For our purposes, we interpret the results of Ref.\cite{Petravic20051} in the following way. 
We assume that if (while) a supercooled liquid remains trapped in an inherent state 
for some (long) time $\tau$ then during this time there is essentially only the vibrational contribution to viscosity.
Transitions between the inherent states represent the contribution to viscosity 
associated with the structural relaxation. 
In this paper we address how the structural relaxation is reflected
in the atomic stress correlation functions. 
The role of the structural relaxation with respect to viscosity has also been 
addressed in recent Ref.\cite{Leporini20121,HarrowellP20121,HarrowellP20161}. 
It is of interest that our present results suggest a picture that is noticeably 
simpler than some of the conclusions presented in Ref.\cite{HarrowellP20161}.
In Ref.\cite{HarrowellP20161} it has been concluded that there is structural 
contribution to viscosity associated with the large distances.
See Fig.6,7,8,9 of Ref.\cite{HarrowellP20161}. 
Our results suggest that one can account for the contribution 
to viscosity from the structural relaxation essentially through the consideration 
of the atomic level stress autocorrelation function, i.e., through the zero-distance contribution. 
This interpretation does not imply that there is only zero distance contribution 
to viscosity from the stress relaxation. 
This result occurs because large distances cause alternating positive and 
negative contributions to viscosity which cancel each other.    

The paper is organized as follows. 
In section \ref{sec:theory} we describe the approach used to study the stress correlations. 
In section \ref{sec:model} we describe the model system that has been used in our simulations. 
In section \ref{sec:results} we describe and discuss the obtained results. 
In the discussion section \ref{sec:discussion} we briefly address the connection between
our results and the results obtained previously within the mode-coupling theory approaches. 
We conclude in section \ref{sec:conclusion}.

\section{Vibrational and Structural Contributions to Viscosity}\label{sec:theory}

\subsection{Expansion of the macroscopic Grenn-Kubo stress correlation function in terms of the atomic level stresses} 

The starting point for the used approach is the well-known Green-Kubo expression for 
(zero-wavevector and zero-frequency) viscosity \cite{HansenJP20061,Green19541,Kubo19571,Helfand19601,Hoheisel19881}:
\begin{eqnarray}
\eta=\frac{V}{k_B T}\int_{0}^{\infty}\left<\sigma^{ab}(t_o)\sigma^{ab}(t_o+t)\right>_{t_o}dt,
\label{eq:green-kubo-01}
\end{eqnarray}
where $V$ is the volume of the system, $k_B$ is the Boltzmann constant, $T$ is the temperature, 
$\sigma^{ab}(t)$ is the $ab$ component of the macroscopic stress tensor of the sample. 
The averaging if done over the equilibrium canonical ensemble 
(in practice, over the initial times, $t_o$, under the assumption that ergodicity holds).

It has been discussed in Ref.\cite{Erpenbeck19951} that the derivations of (\ref{eq:green-kubo-01}) have 
been done for the systems of infinite size without any assumptions with respect
to the applied periodic boundary conditions. However, Eq.\ref{eq:green-kubo-01} is 
routinely used in MD simulations
with the Periodic Boundary Conditions (PBC) applied. 
It follows from Ref.\cite{Erpenbeck19951} and also from our previous investigations 
that the effects of the PBC are far from
being trivial \cite{Levashov20111,Levashov2013,Levashov20141,Levashov2014B}. 
This issue will arise again in this paper.

Further we assume that interactions between the particles can be described by pair potentials.
We also assume that contributions to the stress tensor from the terms associated with the particles' 
velocities can be neglected, as they are usually small in dense supercooled liquids 
in comparison to the contributions from the terms associated with
interactions between the particles \cite{Hoheisel19881}.
In this case, the expression for the macroscopic stress tensor, $\sigma^{ab}(t)$, 
can be written as: \cite{HarrowellP20121,IlgP20151,HarrowellP20161,McQuarrie19761,HansenJP20061,Green19541,Kubo19571,Helfand19601,Hoheisel19881,
Woodcock19911,Woodcock20061,Levashov20111,Levashov2013}:
\begin{eqnarray}
\sigma^{ab}(t)= \frac{\rho_o}{N}\sum_{i=1}^{i=N}s_i^{ab}(t),\;\;\;\;\;\;s_i^{ab}=\sum_{j \neq i} 
\varphi'(r_{ij})\frac{r_{ij}^a r_{ij}^b}{r_{ij}},\;\;\;
\label{eq:green-kubo-02}
\end{eqnarray}
where $\rho_o$ is the number density of the particles,
$r_{ij}=|\vec{r}_j - \vec{r}_i|$ and $\varphi'(r_{ij})=d\varphi(r_{ij})/dr_{ij}$.
In the following, we refer to the quantities $s_i^{ab}$ as to the local atomic stress elements.
We note that the local atomic stress elements differ from the local atomic stresses 
because there is no local atomic volume present in their 
definitions \cite{Egami19801,Egami19802,Egami19821,Chen19881}.
 
Using the first expression from (\ref{eq:green-kubo-02}), expression (\ref{eq:green-kubo-01}) can be rewritten as:
\begin{eqnarray}
\eta=\eta_{auto}+\eta_{cross},
\label{eq:green-kubo-03}
\end{eqnarray}
where
\begin{eqnarray}
&&\eta_{auto}=\frac{\rho_o}{k_B T}\int_{0}^{\infty}F_{auto}(t)dt,\label{eq:green-kubo-04-1}\\
&&\eta_{cross} = \frac{\rho_o}{k_B T}\int_{0}^{\infty}\int_{0}^{\infty}F_{cross}(t,r)drdt,\label{eq:green-kubo-04-2}
\end{eqnarray}
and
\begin{eqnarray}
&&F_{auto}(t) \equiv \left<s^{ab}_i(t_o)s^{ab}_i(t_o+t)\right>_{i,t_o},\label{eq:F-auto-01}\\
&&F_{cross}(t,r) \equiv \left<s^{ab}_i(t_o)\sum_{j\neq i}s^{ab}_j(t_o+t)\delta\left(r-r_{ij}(t_o)\right)\right>_{i,t_o}\label{eq:F-cross-01}
\end{eqnarray}
In (\ref{eq:F-cross-01}) $r_{ij}(t_o)=|\vec{r}_j(t_o) - \vec{r}_i(t_o)|$ and $\delta(r)$-is the delta function.
In treating the data from computer simulations, summation in (\ref{eq:F-cross-01}) 
is assumed to be over those particles $j$ that are within some narrow
interval of distances $[r;r + dr]$ 
from the position of $i$ at time $t_o$.

As far as we know, the expansion of (\ref{eq:green-kubo-01}) into 
expressions (\ref{eq:green-kubo-04-1},\ref{eq:green-kubo-04-2},\ref{eq:F-auto-01},\ref{eq:F-cross-01}) at first has been done in
Ref. \cite{Woodcock19911,Woodcock20061}. 
In our view, this expansion provides a natural view on the atomic scale stress correlations that
determine viscosity. 
Correlations between the atomic scale stresses have been considered 
by several authors in the contexts of the stress relaxation and angular dependent stress fields. \cite{HarrowellP20161,Voigtmann2016,Leporini20121,Lemaitre20151,Chattorai20131,
Puosi20161,Levashov20111,Levashov2013,Levashov20141,Levashov2014B,Levashov20162}.

In the calculations the results of which we present in the following 
we used a further development 
in the interpretations of Eq.(\ref{eq:F-auto-01},\ref{eq:F-cross-01}).
In particular, in order to calculate the stress correlation functions 
in Eq.(\ref{eq:F-auto-01},\ref{eq:F-cross-01}) we used expressions
(51,52,53,54) from Ref. \cite{Levashov20162}. 
This approach automatically performs the averaging over the equivalent stress correlation functions, such as
$\left<s^{xy}_i s^{xy}_j\right>$, $\left<s^{xz}_i s^{xz}_j\right>$, and  $\left<s^{yz}_i s^{yz}_j\right>$.
In any case, for the purposes of this paper, one can simply assume that calculations have been done according 
to the Eq.(\ref{eq:F-auto-01},\ref{eq:F-cross-01}).

\subsection{Correlations between the atomic level pressure elements}

Besides considering correlations between the atomic level shear stress elements, we also present results for the correlations between the atomic level pressure elements. 
The atomic pressure element on every particle is defined as:
\begin{eqnarray}
p_i(t) = \tfrac{1}{3}\left[\sigma_i^{xx}(t)+\sigma_i^{yy}(t)+\sigma_i^{zz}(t)\right].
\label{eq:pi-01}
\end{eqnarray}
The average value of the atomic pressure, in general, is not equal to zero and can be calculated as:
\begin{eqnarray}
\bar{p}_{X}(t) = \tfrac{1}{N}\sum_{i=1}^{i=N}p_{X,i}(t),
\label{eq:pave-01}
\end{eqnarray}
where the ``X" subscript is ``A" or ``B" depending on the type of the particle $i$.
We calculated the average values of the pressure for the particles of type ``A" and the particles of type ``B" separately.
It is known that the average values of the pressure elements on the particles of 
different types are different \cite{Levashov20162,Aur19871}. 
If the average value of the atomic scale pressure is averaged over all particles 
in the system, irrespective of their types, 
then this value corresponds to the macroscopic value of the pressure.
In performing the calculations, we calculated the average value of the pressure 
for every time frame, i.e., we did not use the value of the pressure averaged over the different time frame. 
Thus there is an additional difference with the calculations done for 
the atomic scale shear stress correlation functions -- in those calculations we assume 
that the average value of every atomic scale shear stress component is zero.
For deeply supercooled liquids this issue is a delicate one because 
deeply supercooled liquids can be trapped for a long time in the states 
with non-zero macroscopic shear stress \cite{HarrowellP20121,IlgP20151,HarrowellP20161,Petravic20051}.

In order to define the pressure-pressure correlation function we introduce the following notation: 
\begin{eqnarray}
\mathcal{P}_{X,i}(t) \equiv \left(p_{X,i}(t)-\bar{p}_{X}(t)\right),
\label{eq:Delta-p-01}
\end{eqnarray}
where the average value $\bar{p}_{X}(t)$ is calculated on the same sample 
at time $t$, i.e.,  on the same sample on which $p_{X,i}(t)$ is calculated.

We calculated the correlation function between the atomic 
level pressure elements according to:
\begin{eqnarray}
\Xi_{X,Y}(t,r) \equiv \left<\mathcal{P}_{X,i}(t_o)\mathcal{P}_{Y,j}(t_o+t)\delta(r-r_{ij}(t_o))\right>_{ij,t_o}.\;\;\;\;\;\;\;\;
\label{eq:ppcorr-01}
\end{eqnarray}
In (\ref{eq:ppcorr-01}) the averaging is done over the pairs of particles 
of the proper types and over the initial, ($t_o$), time frames.

\subsection{Correlations between the stress changes of different particles}

Before proceeding further, it is worth noting that the stress correlation 
functions introduced above are closely related to the
correlation functions between the particles' stress changes. 
This observation has also been pointed out in Ref.\cite{Lemaitre20151}.
Indeed, let us consider the following correlation function:
\begin{eqnarray}
\mathcal{D}(t,r) = \left<\Delta s^{ab}_{i}(t_o,t) \cdot \Delta s^{ab}_{j}(t_o,t)\delta(r-r_{ij}(t_o))\right>_{ij,t_o},\;\;\;\;\;\;\;\;\label{eq:D-01}
\end{eqnarray}
where
\begin{eqnarray}
\Delta s^{ab}_{i}(t_o,t) = s^{ab}_{i}(t_o+t)-s^{ab}_{i}(t_o).\;\;\;\label{eq:D-02}
\end{eqnarray}
Thus correlation function (\ref{eq:D-01}) addresses correlations in 
the changes of stresses of particles separated at time $t_o$ by distance $r$.
The expression (\ref{eq:D-01}) can be rewritten as: 
\begin{eqnarray}
\mathcal{D}(t,r) = \mathcal{D}_1(t,r) + \mathcal{D}_2(t,r)  - 2\mathcal{D}_3(t,r),\;\;\;\label{eq:D-03}
\end{eqnarray}
where
\begin{eqnarray}
&&\mathcal{D}_1(t,r) = \left<s^{ab}_{i}(t_o)s^{ab}_{j}(t_o)\delta(r-r_{ij}(t_o)\right>_{ij,t_o},\label{eq:D-04-1}\\
&&\mathcal{D}_2(t,r) = \left<s^{ab}_{i}(t_o+t)s^{ab}_{j}(t_o+t)\delta(r-r_{ij}(t_o)\right>_{ij,t_o},\;\;\;\;\;\label{eq:D-04-2}\\
&&\mathcal{D}_3(t,r) = \left<s^{ab}_{i}(t_o)s^{ab}_{j}(t_o+t)\delta(r-r_{ij}(t_o)\right>_{ij,t_o},\label{eq:D-04-3}
\end{eqnarray}
Note in (\ref{eq:D-04-2}) that the values of the stresses are evaluated 
at time $(t_o+t)$, while the $\delta$-function is calculated at time
$t_o$.
Let us assume that we consider a deeply supercooled liquid and such times $t$ during 
which the positions of the particles do not change significantly (no particles' jumps occur). 
In this situation it is reasonable to expect that: 
$\mathcal{D}_1(t,r) \approx \mathcal{D}_2(t,r)$. 
Under this assumption, it follows from Eq.\ref{eq:F-cross-01} that if $i\neq j$ then 
it is reasonable to expect that:
\begin{eqnarray}
\mathcal{D}(t,r) \sim \Delta\mathcal{F}_{cross}(t,r) \equiv F_{cross}(0,r) - F_{cross}(t,r).\;\;\;\;\;\;\label{eq:D-05}
\end{eqnarray}

\subsection{Structural and vibrational contributions to the particles' stress elements and to the stress correlation function.}

It has been discussed in Ref.\cite{HarrowellP20121,IlgP20151,HarrowellP20161} 
that the inherent states, obtained from the parent liquid at some temperature through 
the steepest descent or conjugate gradient relaxations, possess non-zero macroscopic stresses. 
The considerations in Ref.\cite{HarrowellP20121,IlgP20151,HarrowellP20161} 
are relevant for the finite size systems under the periodic boundary conditions.
However, the Green-Kubo expression for viscosity has been derived under the assumption 
that the system is of infinite (macroscopic) size. 
This issue has been described in some details in Ref.\cite{Erpenbeck19951}. 
Actually, in our view, in the originally derived expression the concept of the 
macroscopic stress is simply absent. 
It is, of course, natural to define the macroscopic stress tensor through the 
atomic scale interactions. However, in our view, the expressions derived originally 
essentially consider the atomic scale stress correlation functions. 
Indeed, according to expressions (\ref{eq:F-auto-01},\ref{eq:F-cross-01}), 
it is reasonable to think about the macroscopic viscosity as about 
the quantity whose value is determined by the microscopic (atomic scale) correlation. 
We adopt this point of view in our considerations.
In this context, we note that it is unclear how to define the macroscopic inherent 
structure stress for an infinite-size system.
For this reason we consider the structural and vibrational contributions 
to the atomic stresses and their correlations without making a reference to 
the macroscopic stresses associated with the inherent structures.

Some additional considerations with respect to the macroscopic 
inherent stresses for the finite size systems are given 
in subsection (\ref{ssec:inherent}).

\subsection{A consideration for the system of infinite size}

Let us assume that we consider a system of infinite size. 
Further we assume that the value of the atomic stress tensor of a particle 
is formed by the structural and vibrational contributions:
\begin{eqnarray}
s^{ab}_{i}(t) = s^{ab}_{str,i}(t)+s^{ab}_{vib,i}(t).
\label{eq:alt-stress-i-01}
\end{eqnarray}
It is possible to think about the structural contribution to the atomic stresses, 
$s^{ab}_{str,i}(t)$, as about the stresses on the atoms in the corresponding inherent structures. 
Using (\ref{eq:green-kubo-01},\ref{eq:green-kubo-02},\ref{eq:alt-stress-i-01}) and 
the assumptions of time-translational invariance and reversibility
we express the macroscopic stress correlation function through the atomic stress correlations:
\begin{eqnarray}
\left(\tfrac{1}{\rho_o^2}\right)&&\left< s^{ab}(t_o)s^{ab}(t_o + t)\right>_{t_o} \label{eq:alt-corrf-04-00}\\
&& = \left<s^{ab}_{str,i}(t_o)\left(\tfrac{1}{N}\right)\sum_{j}s^{ab}_{str,j}(t_o+t)\right>_{ij,t_o}\label{eq:alt-corrf-04-01}\\
&& + 2\left<s^{ab}_{str,i}(t_o)\left(\tfrac{1}{N}\right)\sum_{j}s^{ab}_{vib,j}(t_o+t)\right>_{ij,t_o}\label{eq:alt-corrf-04-02}\\
&& + \left<s^{ab}_{vib,i}(t_o)\left(\tfrac{1}{N}\right)\sum_{ij}s^{ab}_{vib,j}(t_o+t)\right>_{ij,t_o}.\label{eq:alt-corrf-04-03}
\end{eqnarray}
In order to proceed further we assume that there are no 
correlations between the structural and vibrational contributions to the atomic stresses. 
Under this assumption the term (\ref{eq:alt-corrf-04-02}) is equal to zero. 

If the term (\ref{eq:alt-corrf-04-02}) is equal to zero then the expression above suggest that 
it can be assumed that there are two contributions to viscosity; one contribution (\ref{eq:alt-corrf-04-01}) 
is associated with the structural correlations (structural relaxation), while 
another (\ref{eq:alt-corrf-04-03}) with the vibrational correlations (vibrational relaxation).

It follows from our previous publications \cite{Levashov20111,Levashov2013,Levashov20141,Levashov2014B} 
and the data presented in this paper that the vibrational 
relaxation can be interpreted as originating from the decay of the propagating stress waves. 
It has also been briefly discussed in Ref.\cite{Levashov20141} that the structural 
stress relaxation should be related to the decay of the van Hove correlation function. 

In the present paper we show that the contribution to viscosity from the structural 
relaxation can be accounted for through the consideration of the atomic stress 
auto-correlation term (\ref{eq:green-kubo-04-1}).
This does not mean that there are no structural contributions to viscosity from the larger distances. 
This situation arises because the long-range structural correlations carry positive 
and negative contributions to viscosity and thus the long-range structural 
contributions to viscosity effectively cancel each other.

\section{The Model}\label{sec:model}

We studied a binary equimolar system of particles in 3D interacting through purely repulsive potential(s):
\begin{eqnarray}
\varphi_{ab}(r)=\epsilon\left(\frac{\sigma_{ab}}{r}\right)^{12},
\label{eq:potentials}
\end{eqnarray}
where the notations ``$a$" and ``$b$" stand for the types of particles: ``$A$" or ``$B$".
The used values of the parameters are: $\sigma_{AA}=1.0$, $\sigma_{BB}=1.2$, 
$\sigma_{AB}=\sigma_{BA}=1.1=(\sigma_{AA}+\sigma_{BB})/2$. 
The large $B$-particles have masses which are two times larger than the masses of the small $A$-particles: $m_B=2m_A$. 
The chosen value of the particles' number density was: $\rho_o=(N_A+N_B)/V = 0.80$.

\begin{figure}
	\includegraphics[angle=0,width=3.3in]{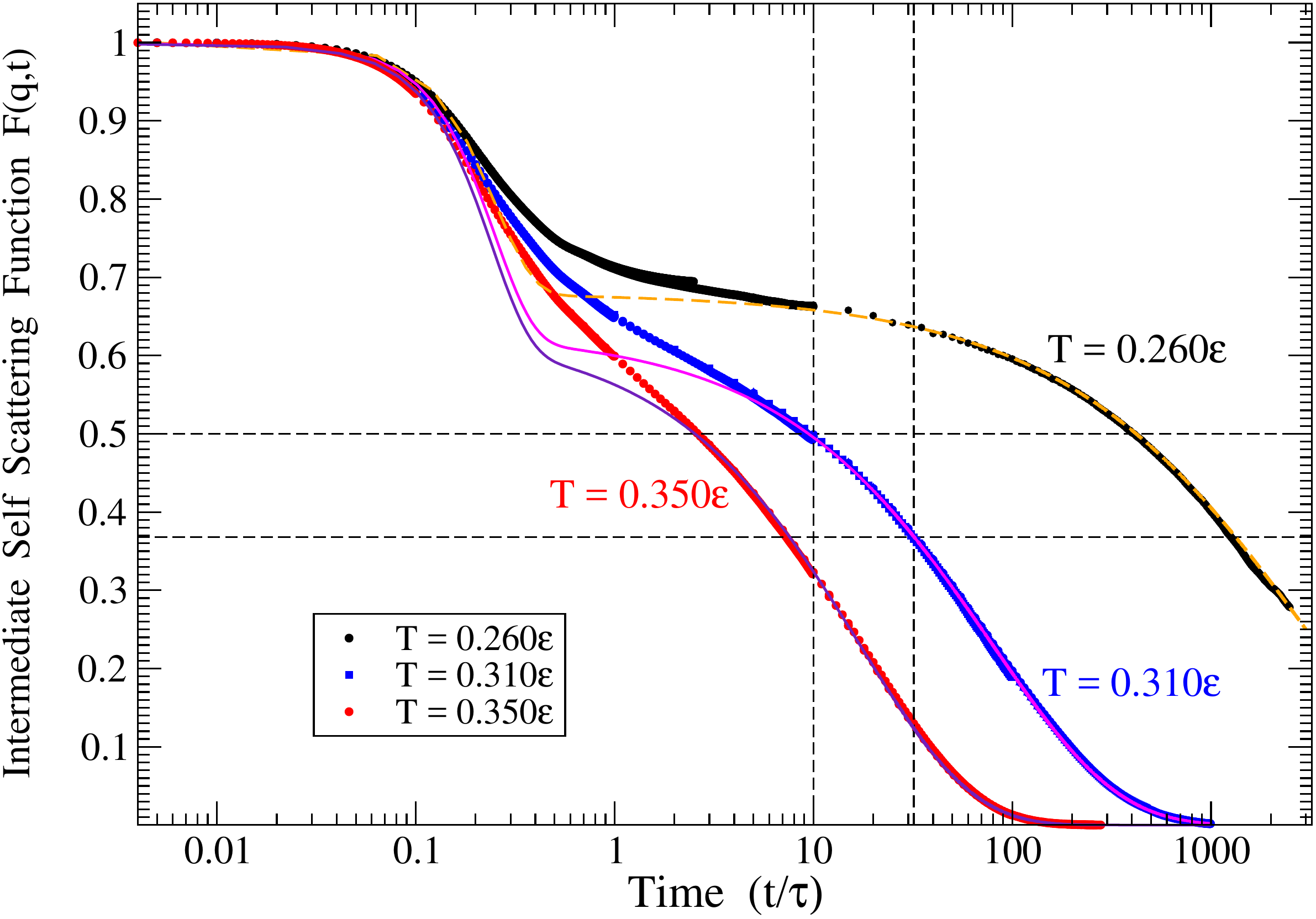}
	\caption{The dependence of the intermediate self-scattering function, $F(q_{max},t)$, 
		on time at temperatures $T=0.350\epsilon$, $T=0.310\epsilon$ and $T=0.260\epsilon$ 
		for the particles of type ``A". The $F(q_{max},t)$ was calculated 
		for the value of $q_{max} =  6.85\sigma^{-1}$ which corresponds to the 
		position of the first maximum in the structure function for the particles of type ``A".
		The dashed vertical lines show the times at which $F(q_{max},t)$ for $T=0.310\epsilon$ decays to $1/2$ and $1/e\approx 0.37$.
		We also attempted to fit the obtained intermediate self-scattering functions with the
		fitting 
		function $F(t)=a \exp\left[-\left(t/\tau_1\right)^{\alpha}\right] + b \exp\left[-\left(t/\tau_2\right)^{\beta}\right]$.
		These fitting functions are shown in the figure with solid lines.
		For $T=0.350\epsilon$ the values of the parameters are as follows: 
		$a=0.37$, $\tau_1 = 0.24\tau$, $\alpha=2.2$, $b=0.63$, $\tau_2 = 17.0\tau$, $\beta = 0.77$.
		For $T=0.310\epsilon$ the values of the parameters are as follows: $a=0.37$, $\tau_1 = 0.25\tau$, $\alpha=2.2$, $b=0.63$, $\tau_2 = 79.0\tau$, $\beta = 0.69$.
		For $T=0.260\epsilon$ the values of the parameters are as follows: $a=0.32$, $\tau_1 = 0.25\tau$, $\alpha=2.2$, $b=0.68$, $\tau_2 = 3000\tau$, $\beta = 0.60$.
	}\label{fig:FqtA-01}
\end{figure}
\begin{figure}
	\includegraphics[angle=0,width=3.3in]{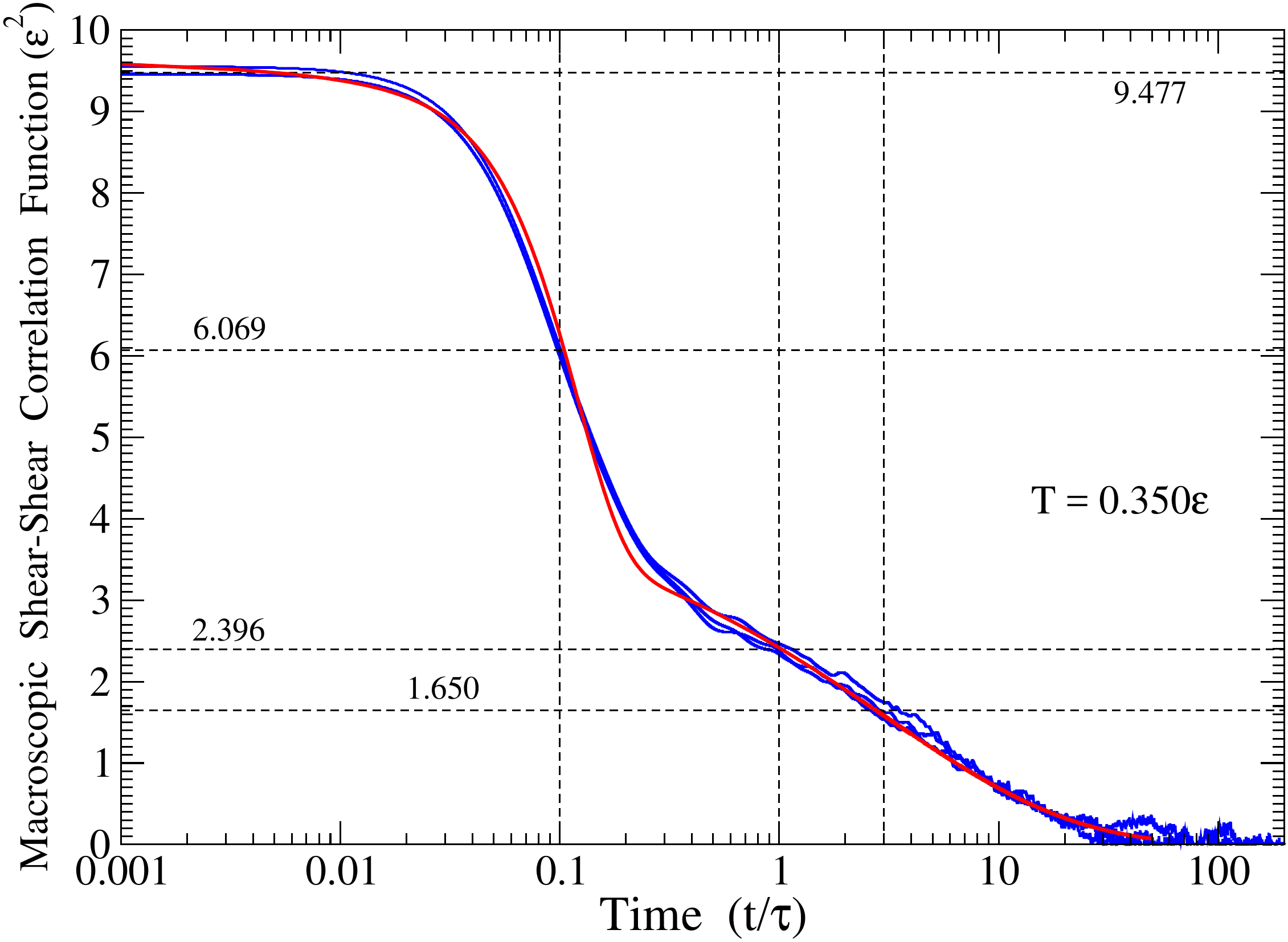}
	\caption{The blue curves show the macroscopic shear stress correlation functions
		$\left<\sigma^{xy}(0)\sigma^{xy}(t)\right>$, 
		$\left<\sigma^{xz}(0)\sigma^{xz}(t)\right>$, 
		$\left<\sigma^{yz}(0)\sigma^{yz}(t)\right>$ 
		calculated for the system of 5324 particles with $(L/2)\approx 9.41\sigma$. 
		The red curve shows the fit of the MD data with the function
		$F(t)=5.36\exp[-(t/(0.12\tau))^{2.0}]+4.3\exp[-(t/(3.0\tau))^{0.5}]$.
		The vertical dashed lines show the times for which we will demonstrate in the following
			the relationship between 
			the microscopic (atomic scale) stress correlation functions and the 
			macroscopic stress correlation function shown in this figure. 
			The horizontal dashed lines are obtained from the intersections of the
			vertical dashed lines with the macroscopic shear stress correlation function.
			The approximate ordinate positions of the intersection points are shown near the corresponding 
			horizontal dashed lines.
	}\label{fig:ss-01-macro}
\end{figure}

In the following we measure the distances in the units of $\sigma_{AA}$ and 
the temperature in the units of $\epsilon$.  
The unit of time that arises from the model's parameters is 
$\tau=\left(\epsilon \sigma_{AA}^2/m_{AA}\right)^2$.
In the following we measure time in the units of $\tau$.

The model that we used in our simulations has been extensively studied previously
\cite{Hansen19881,Miyagawa19911,Mizuno20101,Mizuno20111,Matsuoka2012,Mizuno20131}.
The freezing point of the corresponding one-component system is approximately
$T_m = 0.772\epsilon$ \cite{Miyagawa19911}. 
At $T=0.253\epsilon$ the system is in a deeply supercooled state \cite{Mizuno20101,Mizuno20111,Matsuoka2012,Mizuno20131}.  
In Ref.\cite{Levashov20162} we studied instantaneous properties of the atomic stresses 
and stress correlations in the described model.

Molecular Dynamics simulations have been performed with the LAMMPS program \cite{Plimpton1995,lammps}.
The typical value of the time step used was $0.001\tau$.
The relaxations of the parent liquid structures into the inherent states have been performed with the conjugate gradient method within the LAMMPS program.

\section{Results}\label{sec:results}

\subsection{The studied temperatures and the relevant timescales} 
 
In this paper we present the results for the temperatures $T=0.350\epsilon$, $T=0.310\epsilon$, and $T=0.260\epsilon$.
The temperature $T=0.350\epsilon$ corresponds to the beginning of the true supercooled liquid regime, 
while the temperature $T=0.260\epsilon$ is well in the supercooled region 
\cite{Hansen19881,Miyagawa19911,Mizuno20101,Mizuno20111,Matsuoka2012,Mizuno20131}. 
In order to provide more intuitive insights into these temperatures we show 
in Fig.\ref{fig:FqtA-01} the intermediate self scattering functions for the small ``A" particles at these temperatures. 
More detailed information concerning the timescales of the studied model can be gained from Ref.\cite{Hansen19881,Miyagawa19911,Mizuno20101,Mizuno20111,Matsuoka2012,Mizuno20131}.

\subsection{The macroscopic shear stress correlation function}
 
The macroscopic shear stress correlation function for $T=0.350\epsilon$ is shown in Fig.\ref{fig:ss-01-macro}.
The comparison of Fig.\ref{fig:ss-01-macro} with Fig.\ref{fig:FqtA-01} demonstrates that the shear 
stress correlation function decays noticeably faster than the intermediate self scattering function.
This result also follows from the comparison of the timescales obtained from the fits of the 
curves in these figures with the compressed and stretched exponential functions. 
Thus, for temperature $T=0.350\epsilon$ the time relevant to the long-time decay of the intermediate 
self-scattering function is $\sim 17\tau$, while the time relevant to the long time behavior 
of the shear stress correlation function is almost 6 times smaller, i.e., $\sim 3\tau$.
Note, however, that the fitting functions used for the two curves are different.
The different time dependencies of the intermediate self-scattering function and 
the macroscopic stress correlation function can be understood on the basis of the mode-coupling theory (MCT) \cite{Balucani19881,Balucani19901,Gotze1998sw}.
From Fig.\ref{fig:ss-01-macro} it is quite clear that the time $\sim 0.3\tau$ corresponds 
to the Einstein's vibrational period \cite{HansenJP20061}.

In the following we are going to show that in the time interval 
between $0.3\tau$ and $8.0\tau$ there is a significant contribution 
to the shear stress correlation function 
associated with the propagating shear waves.

\subsection{The microscopic (atomic scale) shear-shear and pressure-pressure correlation functions} 

For the calculations of the stress correlation functions, we had chosen the value of the
bin size $\Delta r = 0.01\sigma$ because it provides good resolution of the 
first peak in the correlation functions. 
For the distances beyond $r=3.0\sigma$, the values of the stress correlation functions for every 
value of $r$ were obtained by averaging over the $11$ values 
of the stress correlation function at the nearest points
(five points from the left, the point itself, and five points from the right). 
This procedure reduces the noise at large distances 
without affecting the actual behavior of the averaged stress correlation curves.
For the distances $r< 3.0\sigma$ the averaging procedure over the $11$ values has not been used.

\begin{figure*}
\begin{center}
\includegraphics[angle=0,width=7.0in]{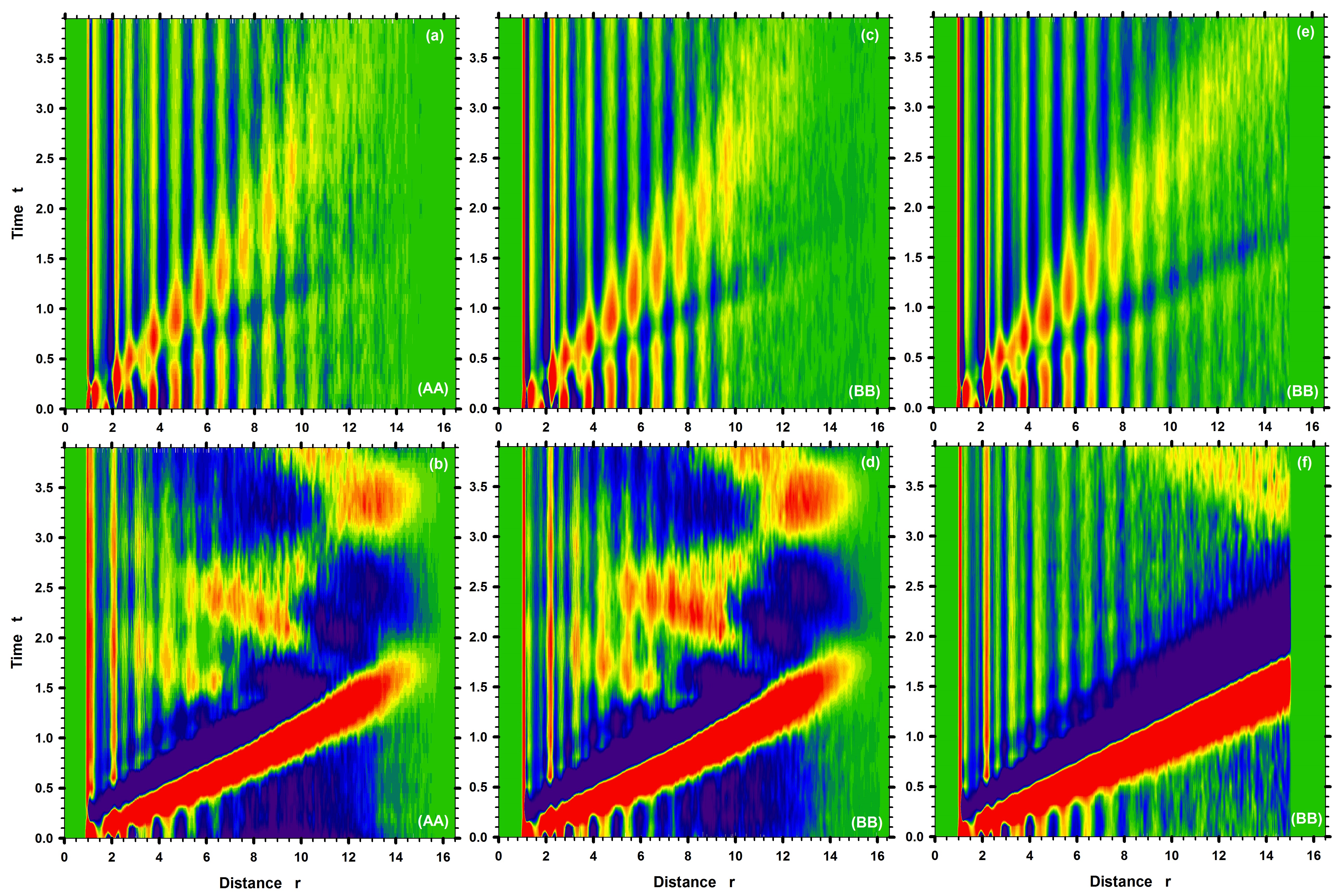}
\caption{The panel $(a)$ shows the normalized \sscf between the particles of 
type ``A" (i.e., ``AA-\sscf"), while the panel $(c)$ shows the normalized \sscf between 
the particles of type ``B" (i.e., ``BB-\sscf"). The distances and times in the
figures are in the units of $\sigma$ and $\tau$.
The calculations have been done according to 
(\ref{eq:F-cross-01}) for the system of $5324$ particles in 
the cubic simulation box with the edges of half-length $(L/2)=9.4\sigma$. 
Thus the half-length of the cube's main diagonal is 
$(D/2) = 16.3\sigma$.
The calculations were done at $T=0.350\epsilon$ which corresponds to 
the beginning of the supercooled region.
The \sscfs have been normalized to the values of the heights of 
the first peaks in the corresponding \sscfs at time zero. 
The scale on the $z$-axis, which is orthogonal to the plane, covers 
the range of intensity $[-0.05;0.05]$ in the $(a)$-panel and the range $[-0.03;0.03]$ in the $(c)$-panel.
The panel $(b)$  shows the normalized pressure-pressure correlation 
function between the particles of 
type ``A" (i.e., ``AA"-\ppcf), 
while the panel $(d)$ shows the normalized ``BB"-\ppcf.
The calculations have been done according to (\ref{eq:ppcorr-01}) 
for the same system of particles and for the same conditions
that have been used to produce panels $(a)$ and $(c)$. 
The scales on the $z$-axis $(b)$ and $(d)$ are the same 
as the scales in $(a)$ and $(c)$ correspondingly.
The panels (e,f) show the \sscf and \ppcf obtained at $T=0.350\epsilon$ 
on the system of $62500$ particles 
with $(L/2) = 21.40\sigma$ and $(D/2) = 37.0\sigma$.
The calculations have been done up to the distance $r=15.0\sigma$. 
The correlations for the particles of types ``BB" are shown.
The scale on the $z$-axis is the same as in panels (c,d), i.e., 
it covers the range of intensity $[-0.03;0.03]$.
It follows from the comparison of the panels [(a) with (c)] and [(b) with (d)] 
that the results for particles of types ``AA" and ``BB" are qualitatively similar.
The results for the ``AB" correlation functions are not shown. 
They look rather similar to the shown correlation functions.
In the upper panels (a,c,e) the major cones of the positive intensity, 
that have their vertexes at the origins, 
correspond to the propagating shear waves.
It is also possible to observe in the panels the footprints of 
the propagating pressure waves--these are less pronounced positive 
intensities that ``propagate" with the higher speed than the cones representing the shear waves.
The pressure-pressure waves shown in panels (b,d,f) support the 
interpretation that in panels (a,c,e) the footprints of the pressure-pressure waves indeed are present.
Note that the pressure waves in panels $(b,d,f)$ are much 
more pronounced than the shear waves in the top panels.
Note also the differences between the panels $(c)$ and $(e)$ and 
the panels $(d)$ and $(f)$. 
It can be seen in panels $(b,d)$ how the pressure-pressure waves 
that leave and reenter the simulation box modify the intensities of the correlation functions.
In the panel $(f)$, obtained on the larger system, 
the pressure-pressure waves, that leave and reenter the simulation 
box, can be observed only for $t > 3.0\tau$ and 
for $r > 10\sigma$.
See the yellow intensity in the upper right corner of panel $(f)$. 
Note also the difference between the panels $(d)$ and $(f)$ for 
the distances larger than $r= 12\sigma$ and for $(t/\tau) \in [1;2]$.
The shear-shear stress waves shown in panels $(a,c,e)$ propagate 
slower, while dissipate faster.
For these reasons the shear-shear waves, that leave and reenter 
 simulation box, can not be clearly seen in the shown panels.
Note, however, the difference between the panels $(c)$ and $(e)$ for 
the distances larger than $r=10\sigma$ and for the times larger than $1.5\tau$.
There are no doubts that these differences are caused 
by the periodic boundary conditions. 
}\label{fig:ss-pp-01}
\end{center}
\end{figure*}

Figure \ref{fig:ss-pp-01} shows the microscopic Shear-Shear Correlation Functions (\sscf) and 
the microscopic Pressure-Pressure Correlation Functions (\ppcf) calculated according to the formulas 
(\ref{eq:green-kubo-04-1},\ref{eq:green-kubo-04-2},\ref{eq:F-auto-01},\ref{eq:F-cross-01})
and (\ref{eq:pi-01},\ref{eq:pave-01},\ref{eq:Delta-p-01},\ref{eq:ppcorr-01}).
In order to obtain these correlation functions the particles coordinates were saved 
with time interval $\Delta t = 0.1\tau$. 
Then the correlation functions for particular times were calculated using the saved particles' coordinates.
Thus, at first, the stress correlations functions for particular times 
were calculated as the functions of distance and then the results for 
the different times we combined to produce Fig.\ref{fig:ss-pp-01}. 
Note that the actual time interval between the time values 
presented in Fig.\ref{fig:ss-pp-01} is $\Delta t = 0.1\tau$.

The similar shear-shear correlation functions have been obtained previously for 
a different single component system of particles in Ref.\cite{Levashov20111,Levashov2013,Levashov20141}. 
The results presented in Fig.\ref{fig:ss-pp-01} establish the generality of the previously obtained results. 
The shown pressure-pressure correlation functions have not been calculated previously.

The results presented Fig.\ref{fig:ss-pp-01} give an impression that the central 
particle at time zero emits the shear and compression waves that propagate away from the central particle as time increases.
This interpretation is reminiscent of the Huygens-Fresnel principle in optics. 
In Ref.\cite{Levashov2014B} we have shown that a picture qualitatively similar 
to the one presented in Fig.\ref{fig:ss-pp-01}
can be obtained through the consideration of the plane 
wave vibrations (phonons) in a model crystal.
Thus, while it is possible to assume that every particle 
in a liquid is a source of the stress waves at every instant,
it is also possible to assume that the stress correlations 
shown in Fig.\ref{fig:ss-pp-01} arise from non-local vibrations. 
\begin{figure}
\begin{center}
\includegraphics[angle=0,width=3.3in]{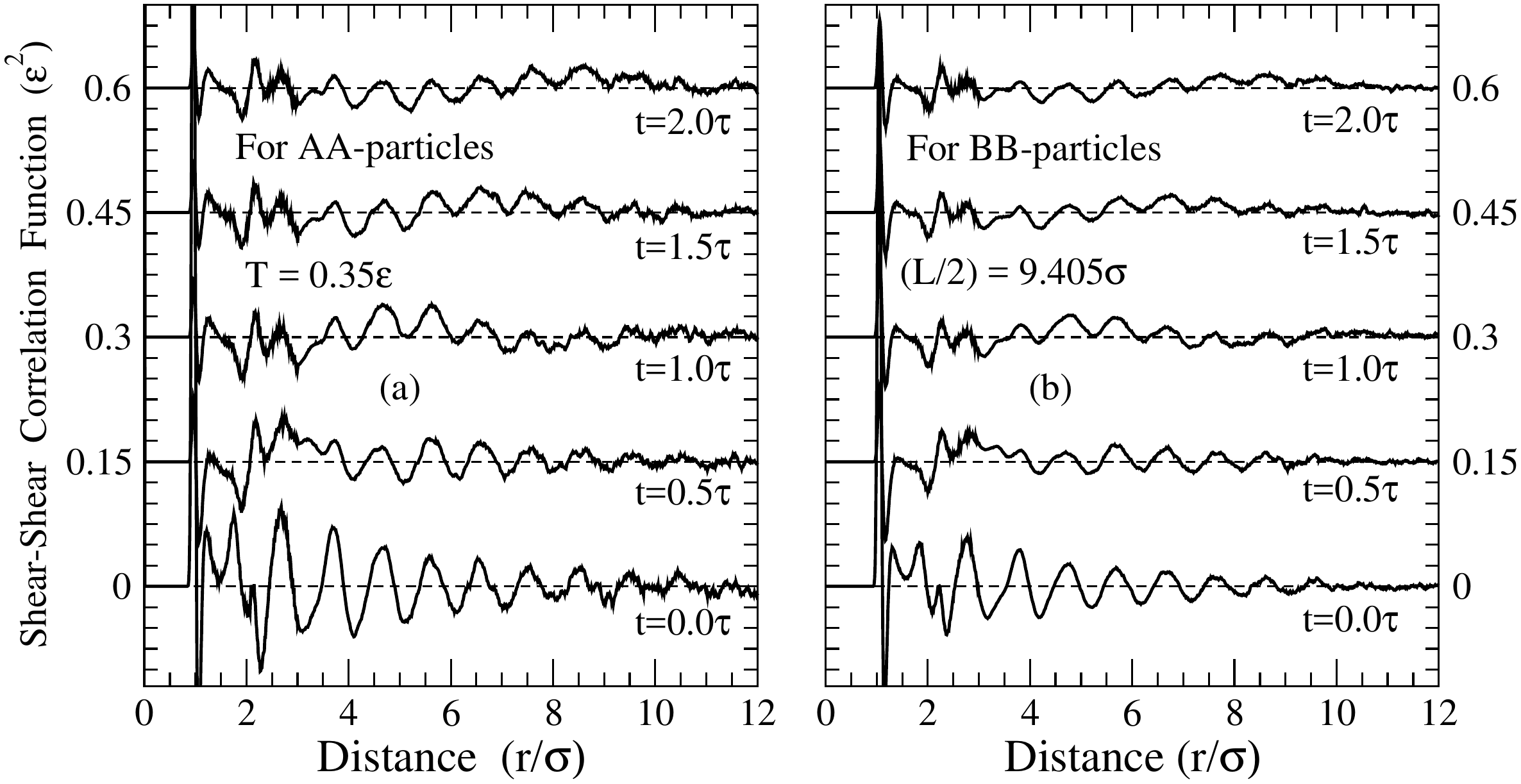}
\caption{The constant time cuts of 
the \sscfs shown in Fig.(\ref{fig:ss-pp-01})(a,c).
The panel (a) shows the constant time cuts of the ``AA"-\sscf, while 
the (b) panel shows the constant time cuts of the ``BB"-\sscf.
The cuts' times are shown near the curves. 
Note the positions of the positive intensities that correspond 
to the propagating shear waves in Fig.(\ref{fig:ss-pp-01})(a,b). 
These intensities can be seen, but overall the shear waves are not very well pronounced.
}\label{fig:ss-cuts-01}
\end{center}
\end{figure}

I order to provide further insight into the data presented in Fig.\ref{fig:ss-pp-01} 
we show in Fig.\ref{fig:ss-cuts-01} the microscopic shear-shear correlation functions 
calculated for some particular times, i.e., for the constant time cuts 
of the stress correlation functions presented in Fig.\ref{fig:ss-pp-01}.
The Fig.\ref{fig:ss-cuts-01}(a) shows the shear-shear correlation 
functions between the particles of type ``A", while
the Fig.\ref{fig:ss-cuts-01}(b) shows the shear-shear correlation 
functions between the particles of type ``B".
Note the positions and the relative intensities of the shear-shear 
waves in the shown curves. 
The curves presented in
Fig.\ref{fig:ss-cuts-01}(a,b) were not normalized to the intensity of 
the first peak associated with the first nearest neighbors, 
while the data in Fig.\ref{fig:ss-pp-01} were normalized. 

\begin{figure}
\begin{center}
\includegraphics[angle=0,width=3.3in]{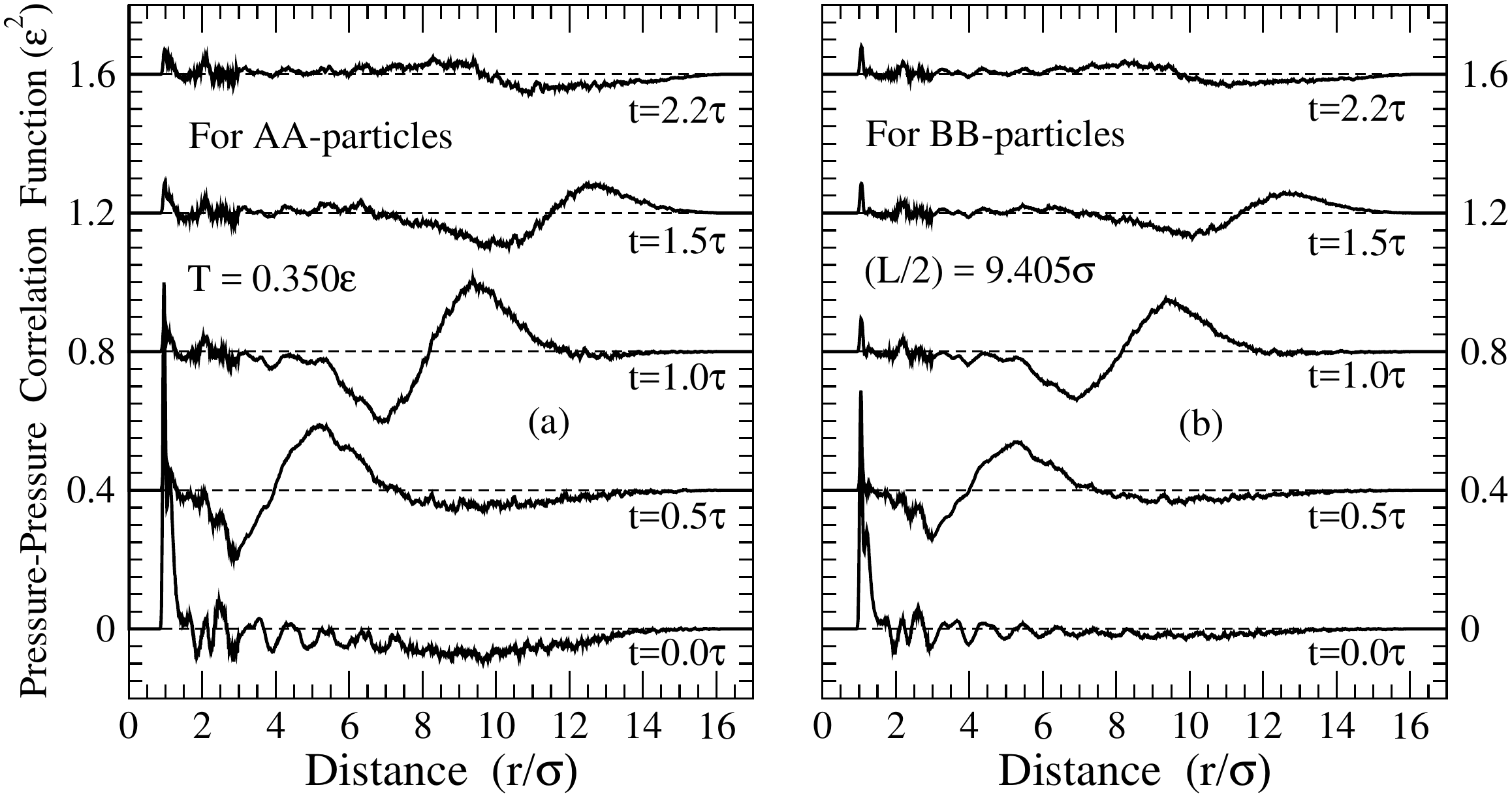}
\caption{The constant time cuts of the \ppcfs shown 
in Fig.(\ref{fig:ss-pp-01})(c,d).
The panel (c) shows the constant time cuts of the ``AA"-\ppcf, while 
the (b) panel shows the constant time cuts of the ``BB"-\ppcf.
The cuts' times are shown near the curves. 
Note that the pressure-pressure waves are much more pronounced than 
the shear-shear waves in Fig.\ref{fig:ss-cuts-01}(a,b).
Note in the curves corresponding to the time $t=2.2\tau$ the 
positive intensity in the interval of distances
$(r/\sigma) \sim [7 ; 10]$. 
This positive intensity is caused by the waves that ``left" 
the simulations box from one side and ``entered" the simulations box from the opposite side. 
See the region of positive intensity located in Fig.(\ref{fig:ss-pp-01})(c,d) in 
the interval of distances $(r/\sigma) \sim [7 ; 10]$ and in the interval of times $(t/\tau) \sim [2.0 ; 2.5]$.
Thus, if one were to integrate over the distances the curves 
corresponding to $t=2.2\tau$ in order to find
their contribution to the bulk viscosity then one would find 
that there is a contribution to the bulk viscosity associated 
with the waves that re-entered the simulation box after leaving it. 
}\label{fig:pp-cuts-01}
\end{center}
\end{figure}
The constant time cuts of the pressure-pressure correlation functions presented 
in Fig.\ref{fig:ss-pp-01}(b,d) 
are shown in Fig.\ref{fig:pp-cuts-01}(a,b). 
From the comparison of Fig.\ref{fig:pp-cuts-01} with Fig.\ref{fig:ss-cuts-01} we 
again can conclude that the pressure-pressure waves are more pronounced than the shear-shear waves. 
Of particular interest is the negative intensity in the pressure-pressure 
correlation function at $t = 0\tau$ in the interval of distances $(r/\sigma) \in [6 ; 13]$.
In our view, the region $r<6\sigma$ can be associated with the pressure 
correlation range for the pressure-pressure correlations. 
Let us now consider the curves corresponding to $t=2.2\tau$. 
These curves exhibit positive intensities in 
the region $(r/\sigma) \in [7 ; 9]$.
From the comparison with Fig.\ref{fig:ss-pp-01}(b,d) it follows that 
this positive intensity arises due to the pressure-pressure waves 
that ``left" and ``reentered" the simulation box.

\subsection{Integral over distance of the microscopic shear-shear correlation function}

\begin{figure}
\begin{center}
\includegraphics[angle=0,width=3.0in]{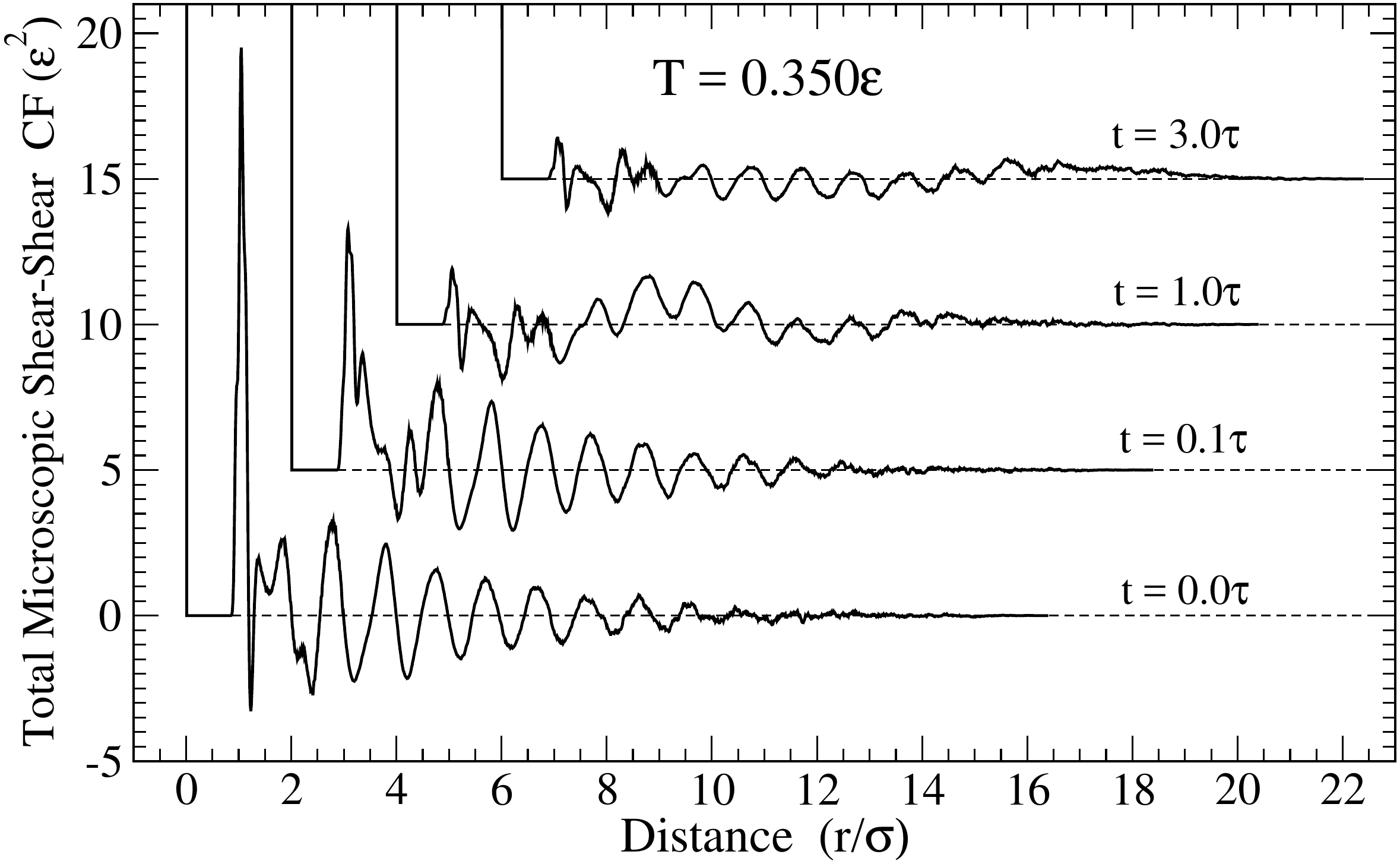}
\caption{The total microscopic \sscfs for the system of 
particles with $(L/2)=9.405\sigma$ at $T= 0.350\epsilon$. 
The shown curves represent the sums of the CFs between the ``AA", ``AB", ``BA", and ``BB" pairs of particles.
The \sscfs between the ``AA" and ``BB" particles are shown in Fig.\ref{fig:ss-cuts-01}.
}\label{fig:micro-shear-shear-cf-01}
\end{center}
\end{figure}

The partial microscopic \sscfs for the ``AA", ``AB", ``BA", and ``BB" particles 
can be summed up to produce the total microscopic \sscf.
These total microscopic \sscfs for the selected times are shown in Fig.\ref{fig:micro-shear-shear-cf-01}.
The selected times in Fig.\ref{fig:micro-shear-shear-cf-01} are 
the same as the selected times in Fig.\ref{fig:ss-01-macro}. 
According to the definition of the microscopic \sscf
(\ref{eq:green-kubo-04-1},\ref{eq:green-kubo-04-2},\ref{eq:F-auto-01},\ref{eq:F-cross-01}), 
the integrals over the distance of the \sscfs shown in Fig.\ref{fig:micro-shear-shear-cf-01}
should be equal to the values of the macroscopic \sscf in Fig.\ref{fig:ss-01-macro} 
at the corresponding times. 

\begin{figure}
\begin{center}
\includegraphics[angle=0,width=3.3in]{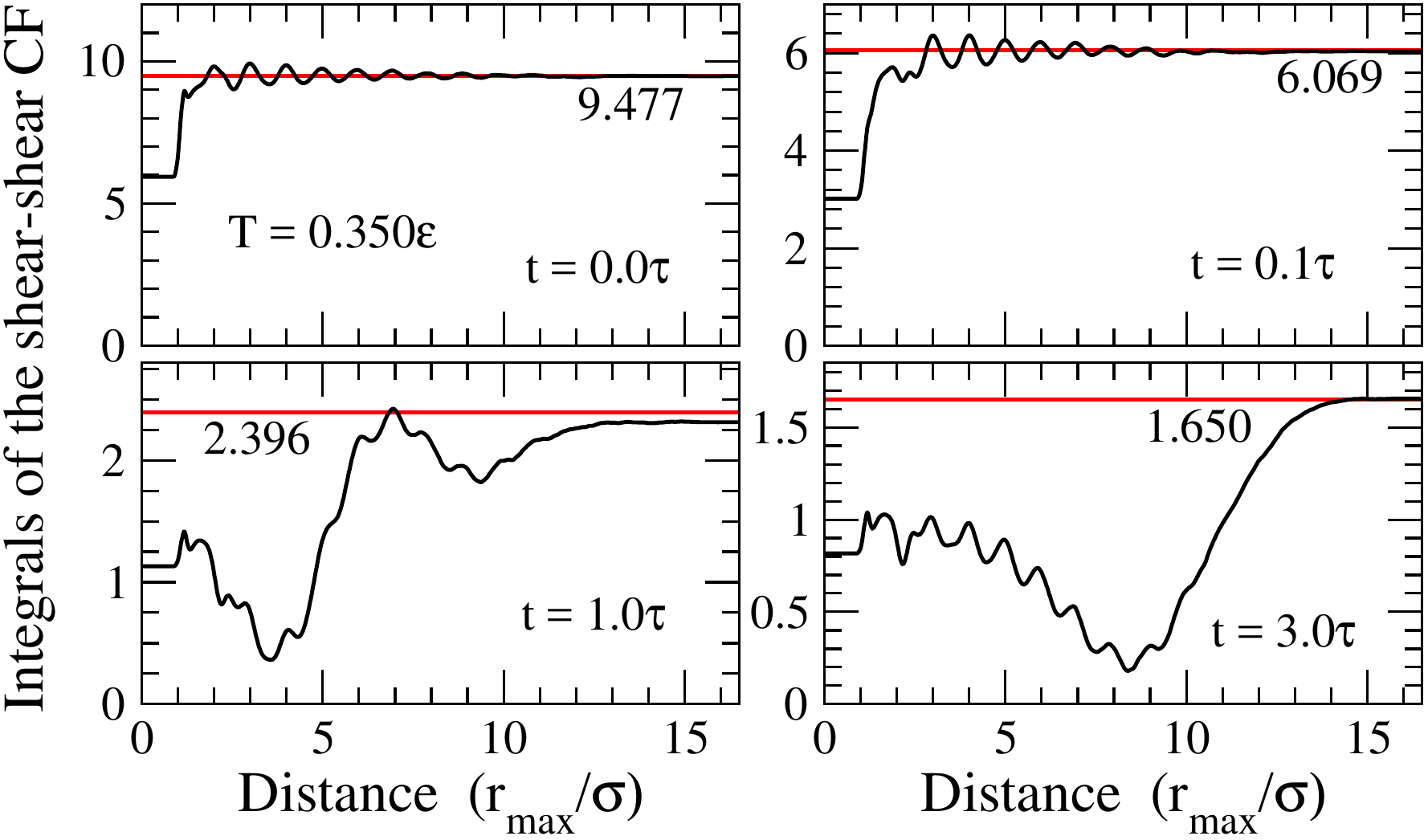}
\caption{The black curves show the integrals over the distance 
of the microscopic \sscfs shown in Fig.\ref{fig:micro-shear-shear-cf-01}. 
The ordinate positions of the red horizontal lines correspond to the values 
of the macroscopic \sscf shown in Fig.\ref{fig:ss-01-macro} at the selected 
times (these ordinate values are shown near the curves at large distances). 
It is clear that the black curves converge to the horizontal red lines 
as all distances become included.
Thus, the figure shows how different distances contribute to the value 
of the macroscopic \sscf at different times.  Note that the curves 
have non-zero values at zero distance. These values originate from 
the shear stress autocorrelation function (\ref{eq:F-auto-01}) which 
has been included (its values can not be properly 
shown in Fig.\ref{fig:micro-shear-shear-cf-01} due 
to their $\delta(r)$-function character). 
}\label{fig:integral-shear-cf-01}
\end{center}
\end{figure}

Figure \ref{fig:integral-shear-cf-01} shows 
how the integrals of the microscopic \sscf (black curves) converge to the macroscopic
\sscf (red horizontal lines) as the upper integration limit, $r_{max}$, increases.
The convergence of the black curves to the red horizontal lines, as all distances become included,
elucidates the role of the shear stress waves and large distances in the formation of viscosity.
 
Note in Fig.\ref{fig:integral-shear-cf-01} that the shown black curves have non-zero values at zero distances.
These non-zero values originate from the stress auto correlation term (\ref{eq:green-kubo-04-1},\ref{eq:F-auto-01}). 
Then at times $t = 0\tau$ and $t = 0.1\tau$ there are rises in the values of the integrals 
associated with the inclusion of the first nearest neighbors. 
It is of interest that at times $t= 1.0\tau$ and $t=3.0\tau$ it is 
less clear if there are contributions associated with the first nearest-neighbor shell. 
In any case, they appear relatively smaller than the contributions 
from the first nearest neighbors at the times $t = 0\tau$ and $t = 0.1\tau$. 
Moreover, it appears that some parts of the contribution associated 
with the first nearest neighbors at small times become associated with the shear waves at later times. 
Note, for the time $t= 1.0\tau$, that the minimum at $r\approx 9.5\sigma$ 
originates from the footprint of the pressure-pressure wave, as 
can be recognized from the behavior of the shear stress 
correlation function in Fig.\ref{fig:ss-pp-01}(a) at $t\approx 1.0\tau$.

\subsection{Structural contribution to the microscopic stress correlation functions}

In our previous considerations of the microscopic \sscfs 
the connection between the propagating shear waves and viscosity has been discussed \cite{Levashov20111,Levashov2013,Levashov20141,Levashov2014B}. 
However, the connection between viscosity and the structural relaxation has not been addressed.
The role of the structural relaxation becomes especially relevant 
in considerations of supercooled liquids.
It has been demonstrated that structural relaxation in supercooled 
liquids proceeds via the localized rearrangement 
events \cite{Picard20041,Leonforte20041,Tanguy20061,Tsamados20081,Leporini20121,Chattorai20131,Chandler20111}. 
Thus there arises a question:
``How these local rearrangement events are expressed in 
the microscopic \sscfs and how do they contribute to viscosity?"

\begin{figure*}
\begin{center}
\includegraphics[angle=0,width=6.6in]{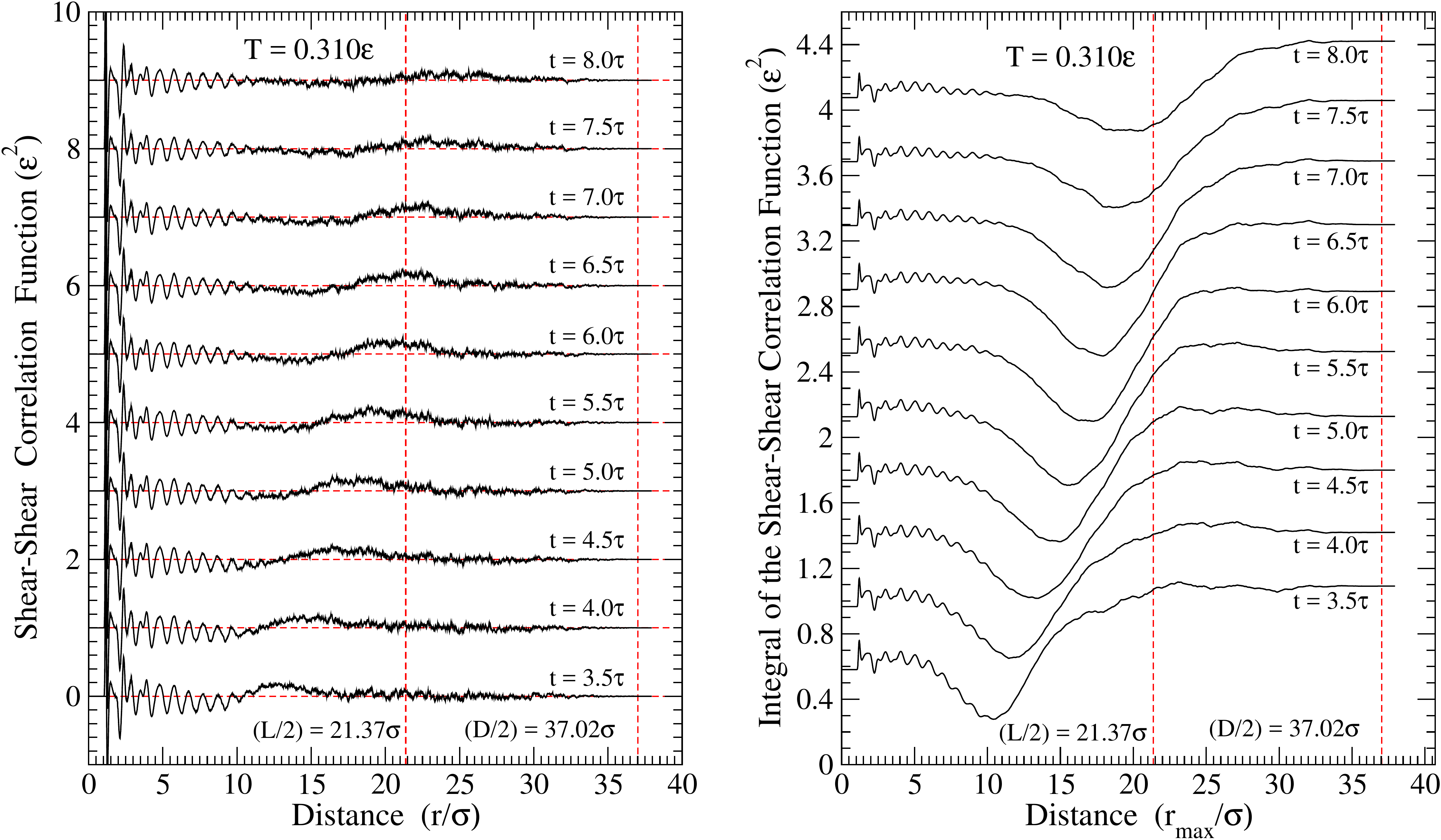}
\caption{The left panel shows the \sscfs for the selected large times calculated 
on the system of 62500 particles at $T=0.310\epsilon$.
For this system the half-length of the cubic simulations box is $(L/2)\approx 21.4\sigma$, 
while the half-length of the main diagonal is $(D/2)\approx 37.0\sigma$. 
The propagating shear stress waves can be clearly observed in the presented curves. 
The purpose of the figure is to address the behavior of the stress correlation function 
in the region of distances which is ``behind" the shear stress wave, i.e., 
in the region of distances between $r=3\sigma$ and $r=10\sigma$.
Note, in particular, the behavior in this region of the curves corresponding to 
$t=6.0\tau$, $t=6.5\tau$, $t=7.0\tau$, $t=7.5\tau$.
The panel on the right shows how the integrals over the distance, $(r/\sigma)$, of 
the curves presented in the left panel depend on the upper integration limit $r_{max}$. 
Thus the curves in the right panel show how the contributions to viscosity 
from some particular times depend on the inclusion distance.
The curves were shifted for the clarity of the presentation. 
The curve corresponding to $t=3.5\tau$ is as it is. 
The curve corresponding to $t=4.0\tau$ was shifted upward by $0.4$. 
The curve corresponding to $t=4.5\tau$ was shifted upward by $0.8$ and so on.
}\label{fig:shear-integral-LL-1}
\end{center}
\end{figure*}

\begin{figure*}
\includegraphics[angle=0,width=6.6in]{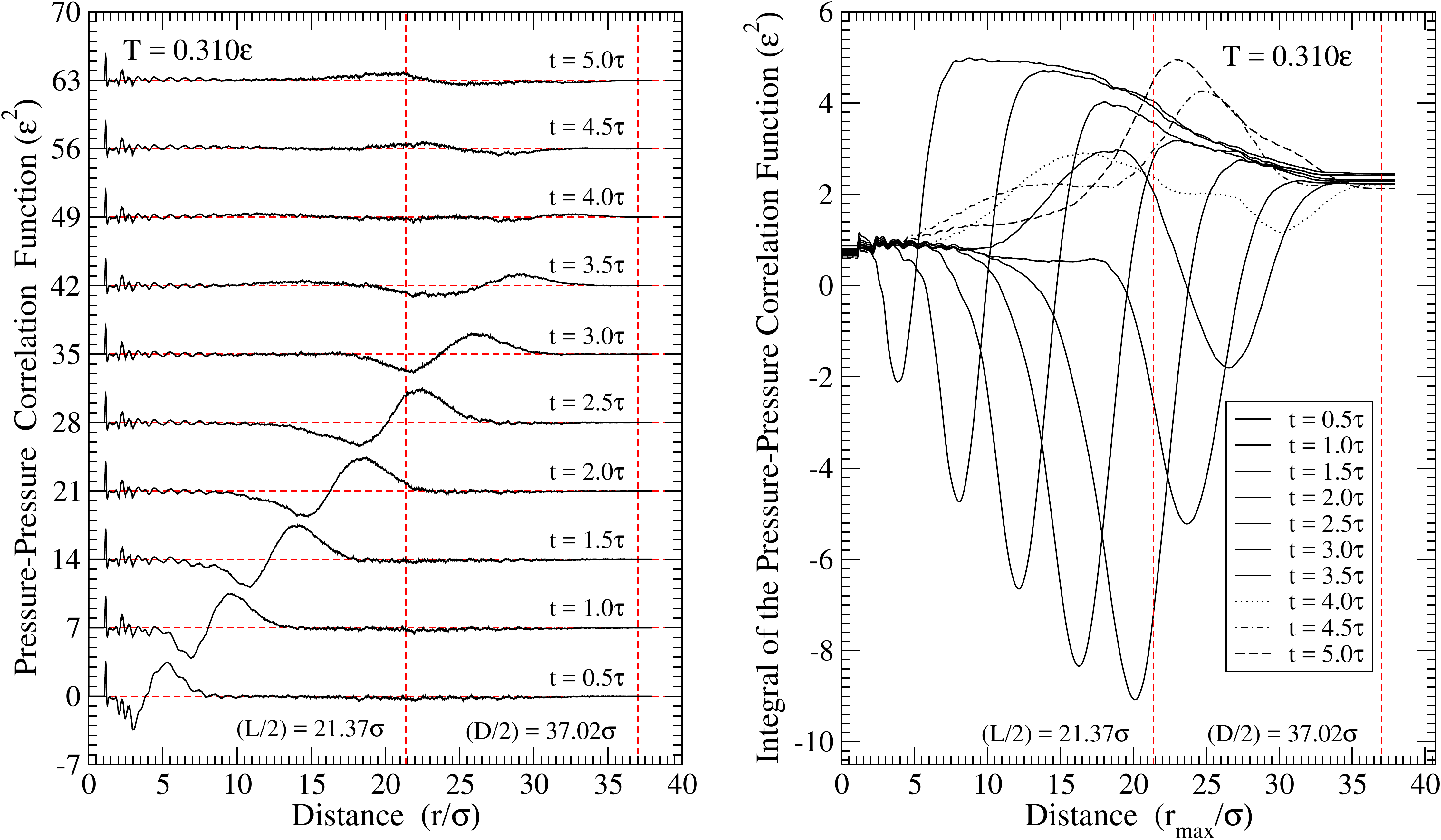}
\caption{The left panel shows the \ppcfs for the selected large times calculated 
on the system of 62500 particles at $T=0.310\epsilon$.
For this system the half-length of the cubic simulations box 
is $(L/2)\approx 21.37\sigma$, while the half-length of the main diagonal is $(D/2)\approx 37.02\sigma$. 
Note that the pressure-pressure wave is more pronounced 
than the shear-shear wave in the left panel of 
Fig.\ref{fig:shear-integral-LL-1}. 
The pressure wave also propagates faster. 
This is in agreement with the results presented in Fig.\ref{fig:pp-cuts-01} for the smaller size system.
For the curve corresponding to $t=3.0\tau$ there is a very well pronounced 
flat region, $(r/\sigma)\approx [3:10]$, behind the pressure wave. 
For the curves corresponding to $t=3.5\tau$ and $t=4.0\tau$ the 
influence of the pressure wave 
that reentered the simulation box can be clearly observed in the \ppcfs.
For $t=3.5\tau$ this reentering makes the pressure-pressure 
correlation curve positive in the region of distances
$(r/\sigma)\approx [11;17]$. 
The panel on the right shows the integrals of the \ppcfs shown in the left panel.
The correspondence between the curves and the times can be 
understood from the positions of the curves' minimums--the positions 
of the curves' minimums increase as time increases.  
Note that all integral-curves have approximately the same value 
at $r=0\sigma$ and approximately the same value after the integration over all distances.
The same value at $r=0\sigma$ means that the contributions 
from the pressure-pressure autocorrelation term are nearly the same for the all presented curves. 
This means that the pressure-pressure autocorrelation term 
remains nearly the same for the all presented times, i.e., 
it means that the pressure-pressure autocorrelation term 
decays very slowly in time. 
This slowness is the expression of the slow structural relaxation.
The fact that all integral curves converge to approximately 
the same value after the integration over all distances means that
the contribution from the pressure-pressure waves is nearly the same for all shown times. 
This, in our view, means that the dissipation of the pressure-pressure waves is slow. 
}\label{fig:pressure-integral-LL-1}
\end{figure*}

In our discussion of Fig.\ref{fig:integral-shear-cf-01} we noticed that there is a contribution 
to viscosity associated with the 
autocorrelation term in the \sscfs. In the following, we suggest that the structural 
contribution to viscosity at large times can be taken into account through 
the consideration of this autocorrelation term. In order to demonstrate this, 
we performed calculations of the microscopic \sscf
for the large times up to the distances that allow to include all particles in the large system.
Thus, we considered the system of $62500$ particles with $(L/2)\approx 21.37\sigma$ (i.e., $(D/2)=37.02\sigma$).
We studied this system at $T=0.310\epsilon$. 
We saved the particles coordinates every $0.5\tau$ and then 
used these coordinates to calculate the shear-shear and the pressure-pressure correlation functions. 
The results for the selected large times are shown 
in Fig.\ref{fig:shear-integral-LL-1},\ref{fig:pressure-integral-LL-1}.

In the left panel of Fig.\ref{fig:shear-integral-LL-1} we see that for the 
time $t=3.5\tau$ the shear stress wave has not yet reached the distance $(L/2)$. 
Thus the positive peak of the shear stress wave is located 
at $r \approx 13\sigma$, 
(``the center" of the wave, according to Ref.\cite{Levashov2014B}, is actually 
located at $r=10.5\sigma$, i.e., at the point where the \sscf crosses the time axis.) 
The curves corresponding to the 
times $t=6.0\,\tau$, $t=6.5\,\tau$, $t=7.0\,\tau$, and $t=7.5\,\tau$ 
represent the situation when some parts of the ``shear waves" have already left the simulation box at $(L/2)$. 
Note, however, that after reentering the simulation box, these shear waves have not yet reached 
the distance $r \approx 10\sigma$.
This can be seen from how the shear stress waves disappear as they move further along the main diagonal beyond $(L/2)$.
Thus, for $t=7.0\tau$ the shear wave has not yet reached the distance $r=30\sigma \approx (21.4+10)\sigma$. 
This also means that the part of the wave that reentered the box has 
not yet reached the distance $r=10\sigma \approx(21.4-10)\sigma$.
Further note that, besides the oscillations, the region of distances between $r=3\sigma$ and $r=10\sigma$ is essentially flat. 
In our view, this region at the times $t=6.0\,\tau$, $t=6.5\,\tau$, $t=7.0\,\tau$, $t=7.5\,\tau$ 
represents the general behavior of the ``behind the wave region" that one should observe in the system of infinite size.
In our view, the presented results suggest that in the system of infinite 
size the "behind the wave region" should be essentially flat, if the oscillations are ignored.  
Note also that the amplitudes and the shapes of the oscillations in 
the region between $r=3\sigma$ and $r=10\sigma$ are quite similar in all curves. 
The change in these oscillations can be considered as an intuitive measure of the structural relaxation. 
It follows, from this perspective, that there occur only some moderate 
structural relaxation as time increases from $t=3.5\,\tau$ to $t=8.0\,\tau$. 

Let us now consider the curves in the right panel of Fig.\ref{fig:shear-integral-LL-1}. 
These curves show how the integrals of the curves from the left panel depend 
on the upper integration limit $r_{max}$.
There are four points with respect to the curves in the right panel that we would like to note.\\ 
1) The values of the curves at $r=0\sigma$ are not zero. 
These non-zero values are due to the contributions from the shear-shear 
autocorrelation term (\ref{eq:F-auto-01}) that has been taken into account.\\ 
2) There is essentially a flat region between $r=3\sigma$ and $r=10\sigma$ for the curves corresponding 
to the times $t=6.0\,\tau$, $t=6.5\,\tau$, $t=7.0\,\tau$, and $t=7.5\,\tau$. 
It is quite clear, in our view, that the small decreases in the values of these integral 
curves around $(r_{max}/\sigma)\approx [7;10]$ are due to the negative tails of the propagating shear waves.\\
3) As the distance increases beyond $r=10\sigma$ there is, at first, the negative contribution 
to integral curves from the ``negative tails" of the shear waves and then there 
is a positive contribution from the ``positive fronts" of the shear waves. 
The overall contributions of the shear waves to the integral curves are positive.\\
4) Finally, note that there are only small changes in the shape and 
the amplitude of the peak originating from the nearest neighbor shell, i.e., in 
the peak located at $(r/\sigma) \in [\approx 1.0 ; \approx 2.5]$ in all curves.
This means that for all considered times the contribution to viscosity 
from the nearest neighbor shell is approximately the same. 
Thus, it is quite natural to expect that for the considered times 
the contributions to viscosity from the autocorrelation 
term (i.e., from the $\delta(r)$-peak) are not very different. 

The left panel of Fig.\ref{fig:pressure-integral-LL-1} shows 
the microscopic \ppcfs at large times.
Note that there, as in the left panel of Fig.\ref{fig:shear-integral-LL-1}, are essentially flat 
behind the wave regions (besides the structural oscillations). 
Thus we come to the conclusion that the structural contribution to the
macroscopic \ppcf (i.e., to the integral over the distance of the microscopic \ppcf)
can be estimated through the consideration of the microscopic pressure-pressure auto-correlation term only. 
Thus the situation with the \ppcf is similar to the situation with the \sscf.
The right panel of Fig.\ref{fig:pressure-integral-LL-1} shows 
the integrals over the distance of the curves shown in the left panel.
This time we did not shift the curves vertically, in contrast 
to the integral curves in Fig.\ref{fig:shear-integral-LL-1}.
Note that all curves have almost the same value at zero distances. 
Thus for the discussed times the pressure-pressure auto-correlation 
term changes only a little.
This means, according to our interpretation of these changes 
as being related to the structural relaxation, 
that structural relaxation proceeds quite slowly.
It is more surprising that the integral curves, after 
the integration over all distances, again converge to almost the same value. 
It means that the vibrational contributions to the integral 
curves at the considered times are somewhat similar.
We speculate on the origin of this behavior from 
the perspective of dissipation of the pressure waves, 
i.e., we assume that the waves' contribution to the integral curves 
is related to the loss of intensity of the pressure waves. 
In our view, it can be assumed that the pressure wave is already ``formed" at $t=0.5\,\tau$. 
Further, it can be assumed that the rate of dissipation of 
the pressure wave so small that the overall intensity of the pressure 
wave does not change significantly over the time interval $\Delta t \approx 5.0\,\tau$. 
On the other hand, it is natural to assume that the rate of the energy loss of 
the pressure wave is proportional to the energy of the pressure wave.
Thus, it appears that the data suggest that the energy of the wave 
for the discusses times decay exponentially in time 
with a rather large dissipation time $\tau$:
$E(t)\sim E_w \exp{[-(t-t_w)/\tau]}$, where $E_w$ is 
the energy of the pressure wave at $t=t_w$.  

The given above description of the obtain data leads us to the following conclusions. 
It is possible to speak about the vibrational and the structural contributions to viscosity. 
The vibrational contribution originates from the propagating shear waves. 
The value of the structural contribution is essentially determined by the contribution that comes from the 
shear stress autocorrelation term (\ref{eq:green-kubo-04-1},\ref{eq:F-auto-01})
(and a small addition from the first peak in the shear-shear correlation function). 
The region of distances beyond the first nearest neighbors contains alternating positive and negative structural contributions to viscosity. 
These positive and negative structural contributions from the large distances mutually cancel each other. 
This cancellation leads to the situation in which the structural contribution to viscosity can be accounted through the consideration of the
stress autocorrelation term and a small contribution associated with the first peak in the shear stress correlation function.

\subsection{Results for larger times}

\begin{figure}
\includegraphics[angle=0,width=3.0in]{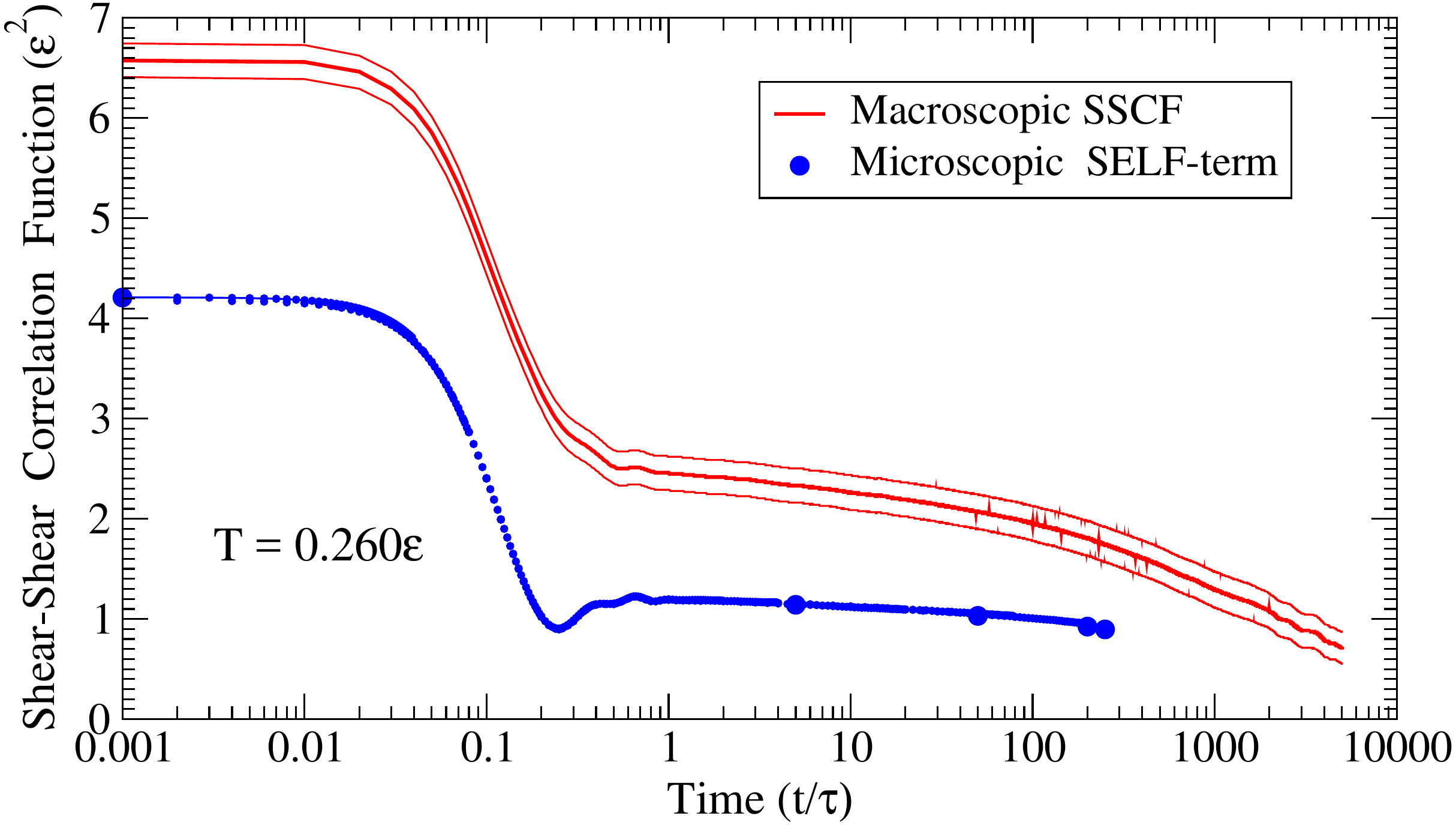}
\caption{The thick red curve is the macroscopic \sscf calculated on the system of $N=5324$ particles
with $(L/2) = 9.405\sigma$ at $T=0.260\epsilon$. 
The two thin red curves show the error of the mean.
The blue circles show the values of the microscopic autocorrelation (or self) term.
The larger blue circles correspond to the times
at which the microscopic \sscfs shown 
in Fig.\ref{fig:large-times-sscfI-1} has been calculated. 
It is of interest to compare this figure with Fig.\ref{fig:ss-01-macro}.
}\label{fig:macro-and-self-1}
\end{figure}

\begin{figure}
\includegraphics[angle=0,width=3.3in]{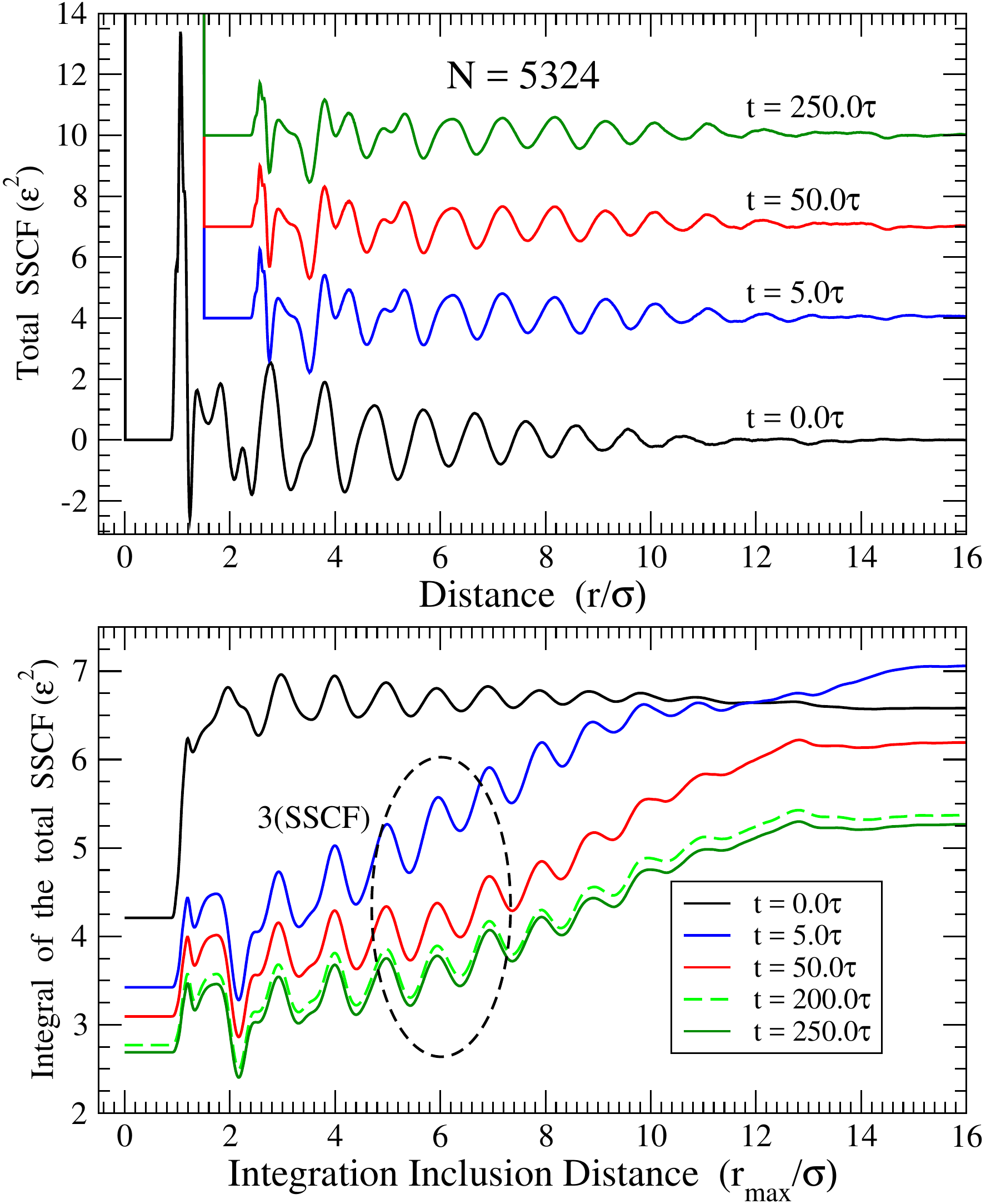}
\caption{The upper panel shows the microscopic \sscfs at the selected times calculated 
on the system of 5324 particles with $(L/2) = 9.405\sigma$ [$(D/2)=16.28\sigma$] at $T=0.260\epsilon$. 
The word ``Total" on the $y$-axes signifies that the shown \sscfs are 
the sums of the partial contributions from the ``AA", ``AB", ``BA", and ``BB" particles. 
Note that the shear waves can not be observed in the shown curves. 
This, however, does not mean that the shear waves do not contribute to 
the integrals of these \sscfs over the distance,
as follows from the bottom panel. Note in the bottom panel that the curve 
corresponding to $t=0\tau$ exhibits the contribution from
the autocorrelation (self) term and also the contribution from the first neighbors. 
Also note that the $t=0\tau$ curve does not exhibit the contribution from the large distances. 
The values of the \sscfs that correspond to $t>0\tau$ 
(the curves which are encircled by the ellipse) were multiplied by 3 (three). 
The presence of this multiplication factor is designated by the ``$3$(\sscf)" mark.  
Note that these curves exhibit the increase at large distances.  
In our view, this increase is caused by the shear waves
which left and reentered the simulation box. 
For the time $t=5\tau$ this reentering happened once, while for the larger times
the reentering happened many times. 
}\label{fig:large-times-sscfI-1}
\end{figure}

In view of the results presented in Fig.\ref{fig:shear-integral-LL-1},\ref{fig:pressure-integral-LL-1},
it is reasonable to consider also the behavior of the microscopic \sscf at the 
times much larger than the time required for the shear waves to propagate through the system.
In particular, compare the contribution to the macroscopic \sscf
from the autocorrelation stress term with the contribution from the propagating shear waves.
The results of the relevant calculations for the system with $(L/2) = 9.405\sigma$ are shown 
in Fig.\ref{fig:macro-and-self-1},\ref{fig:large-times-sscfI-1}(a,b) for $T=0.260\epsilon$ 
which is deeper in the supercooled region than the temperatures considered before.

The thick red curve in Fig.\ref{fig:macro-and-self-1} shows the dependence of the macroscopic \sscf on time for $T=0.260\epsilon$.
The blue curve in Fig.\ref{fig:macro-and-self-1} shows the contribution 
to the macroscopic \sscf associated with the autocorrelation self-term. 
It is natural to associate the difference between the red and blue curves with
the non-local contribution which should be related to the shear stress waves. 

Figure Fig.\ref{fig:large-times-sscfI-1}(a) shows the total microscopic shear stress correlation
functions at the selected times, while Fig.\ref{fig:large-times-sscfI-1}(b) shows
the integrals over the distance of the curves shown in Fig.\ref{fig:large-times-sscfI-1}(a).

In agreement with Fig.\ref{fig:macro-and-self-1}, it follows from Fig.\ref{fig:large-times-sscfI-1}(b)
that at large times there is the contribution to viscosity 
associated with the distances beyond the first neighbors. 

In view of the results presented before, it is quite natural to
associate this long-range contribution with the shear stress waves that passed many times through the system and which are
essentially distributed over the whole finite-size simulation box. 
In other words, it means that the positive and homogeneous contribution from the large distances 
is a finite size effect, i.e., if the system were of infinite size then there would be 
instead the contribution from the very large distances corresponding to the position of the shear wave.

\subsection{Microscopic \sscf from the inherent structures}\label{ssec:inherent}

In Ref.\cite{HarrowellP20161} the microscopic \sscfs calculated on the inherent 
structures have been considered. 
It is natural to expect that relaxation into the inherent states 
removes the vibrational contributions to the atomic stresses 
and, as a consequence, to the \sscfs and to the integrals over 
the stress correlation functions. 
However, the results obtained in Ref.\cite{HarrowellP20161} show that there 
is a positive contribution to the integral of the \sscf 
associated with the distances beyond the first neighbors. 

\begin{figure}
\includegraphics[angle=0,width=3.0in]{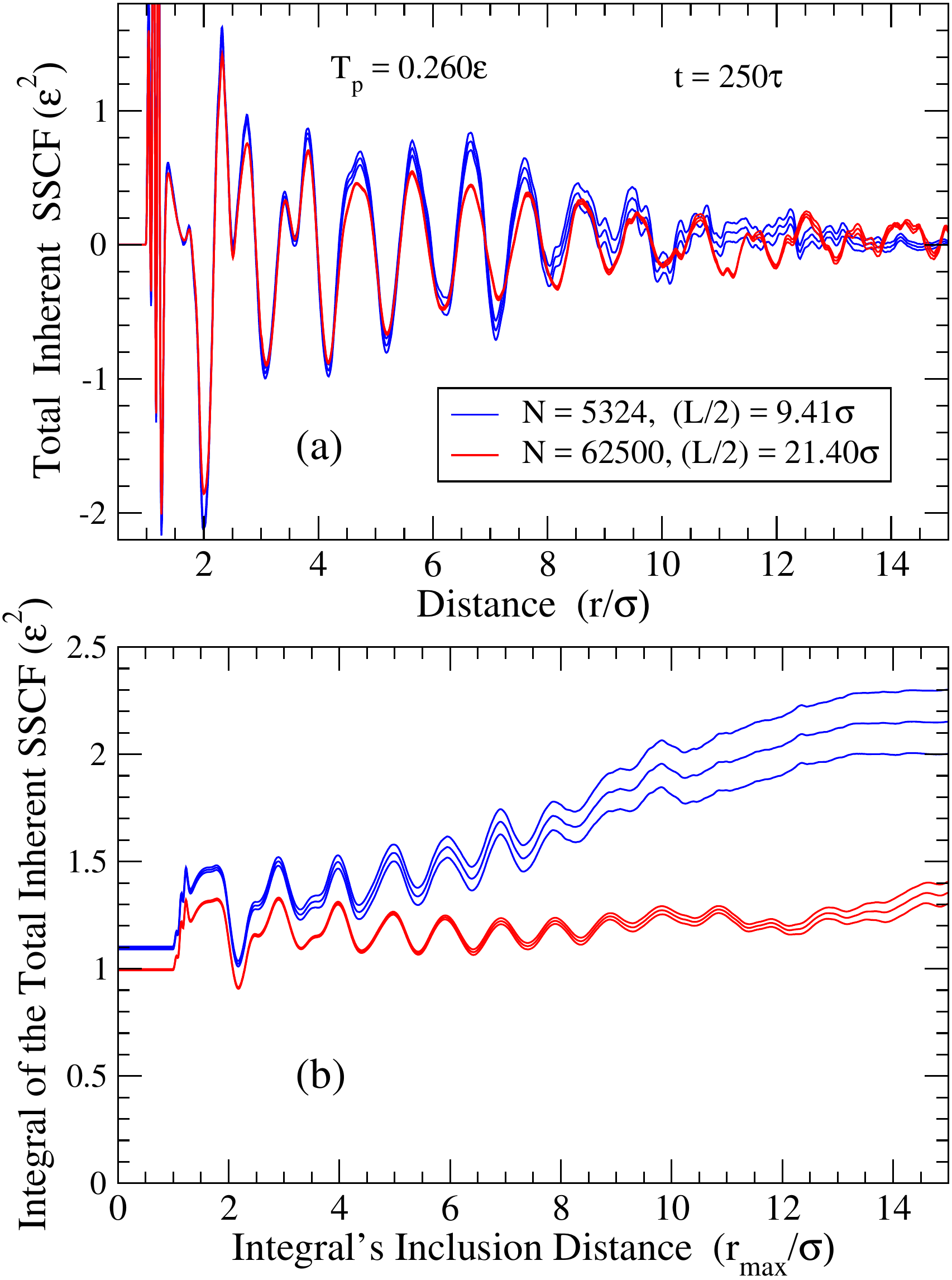}
\caption{Panel (a) shows the total \sscfs calculated on the inherent structures
separated by $250.0\tau$. There are three blue curves that were obtained by averaging 
over 54 independent runs on the systems with 5324 particles. 
There were 3000 initial structures in every run. 
One blue curve corresponds to the average value, while the other 
two represent the average value plus/minus the error of the mean. 
The red curves were obtained on the systems with 62500 particles 
by averaging over 5 independent runs with 1000 initial structures in each run.
Panel (b): The integrals over the distance of the total \sscfs obtained 
in the individual (MD plus Relaxation) runs can be calculated.
The middle blue and the middle red curves represent the averaged 
values of the integrals of the proper \sscfs.
The other two blue and two red curves show the averaged 
values of the integrals plus/minus the proper errors of the mean.
It is clear that the averaged integrals exhibit the dependence on the system's size.
}\label{fig:large-times-inher-1}
\end{figure}

To understand the origin of this situation we also produced the inherent
structures from the parent liquid at $T=0.260\epsilon$ and calculated the 
microscopic \sscfs on these structures.
The inherent structures were produced by the conjugate gradient minimization 
procedure within the LAMMPS program. 
We performed calculations of the inherent microscopic \sscfs on the systems 
with $(L/2) = 9.405\epsilon$ and $(L/2)= 21.37\epsilon$. 
Figure \ref{fig:large-times-inher-1}(a) shows that the results for the \sscfs obtained on two systems 
look rather similar. 
However, the small differences between the \sscfs lead to the noticeable
differences between the integrals of these \sscfs over the distances, 
as Fig.\ref{fig:large-times-inher-1}(b) shows.

In our view, the different behaviors of the integrals of the \sscfs 
at large distances are caused by the different values of 
the fluctuations of the macroscopic inherent structure stresses, 
as discussed below.

At first, we introduce several notations. 
We assume that the atomic level stress element of a particle $i$ in 
an inherent configuration, $s_{inh,i}^{ab}$, can be decomposed into the two parts:
\begin{eqnarray}
s^{ab}_{inh,i}(t) = \bar{s}^{ab}_{inh}(t)+s^{ab}_{str,i}(t),
\label{eq:separation-01}
\end{eqnarray}
where the macroscopic inherent structure stress, $\bar{s}^{ab}_{inh}(t)$,  
and the atomic structural stress, $s^{ab}_{str,i}(t)$, are defined as: 
\begin{eqnarray}
&&\bar{s}^{ab}_{inh}(t) \equiv \sigma^{ab}_{inh}(t) \equiv \frac{1}{N}\sum_{i}s^{ab}_{inh,i}(t),\label{eq:separation-02-1}\\
&&s^{ab}_{str,i}(t) \equiv s^{ab}_{inh,i}(t) -\bar{s}^{ab}_{inh}(t).\label{eq:separation-02-2}\;\;\;
\label{eq:separation-02}
\end{eqnarray}
According to Ref.\cite{HarrowellP20121,IlgP20151,HarrowellP20161}, the macroscopic 
inherent stress, $\bar{s}^{ab}_{inh}(t)=\sigma^{ab}_{inh}(t)$, in general, is not equal to zero.

In calculations of the microscopic stress correlation functions we consider the correlations
between the inherent stress elements of particle $i$, $s_{inh,i}^{ab}$, with 
the inherent stress elements of those particles $j$, $s_{inh,j}^{ab}$, which are distance $r$ 
away from the particle $i$ at time $t_o$:
\begin{eqnarray}
&&\left<s^{ab}_{inh,i}(t_o)\sum_{j}s^{ab}_{inh,j}(t_o + t)\delta(r-r_{ij}(t_o))\right> \label{eq:corrf-03-00}\\
&& = \left<\bar{s}^{ab}_{inh}(t_o)\bar{s}^{ab}_{inh}(t_o+t)\sum_{j}\delta(r-r_{ij}(t_o))\right>\label{eq:corrf-03-01}\\
&& + \left<\bar{s}^{ab}_{inh}(t_o)\sum_{j}s^{ab}_{str,j}(t_o+t)\delta(r-r_{ij}(t_o))\right>\label{eq:corrf-03-02}\\
&& + \left<s^{ab}_{str,i}(t_o)\bar{s}^{ab}_{inh}(t_o+t)\sum_{j}\delta(r-r_{ij}(t_o))\right>\label{eq:corrf-03-04}\\
&& + \left<s^{ab}_{str,i}\sum_{ij}(t_o)s^{ab}_{str,j}(t_o+t)\delta(r-r_{ij}(t_o))\right>\label{eq:corrf-03-06}
\end{eqnarray}
The expression above can be simplified if several assumptions are made. 
In particular, in our view, it is reasonable to assume that there is no correlation between the
structural atomic stress and the average macroscopic inherent stress. 
This is essentially the consequence of 
the definitions (\ref{eq:separation-02-1},\ref{eq:separation-02-2}).
Thus (\ref{eq:corrf-03-02}) and (\ref{eq:corrf-03-04}) are equal to zero. 
It is also reasonable to assume that there is a positive
correlation between the average inherent stresses at times $t_o$ and $t_o + t$.
Under this assumption the contribution from (\ref{eq:corrf-03-01}) grows 
with the increase of\, $r$. 
The smaller is the size of the system the larger are fluctuations in the
values of the average inherent stresses. 
Thus, it is naturally to expect that the value of the term (\ref{eq:corrf-03-01}) 
increases as system's size decreases. 
In our view, this might be the explanation of 
the large-distance contribution to the integral of the microscopic 
stress correlation function calculated on the inherent structures.

\section{Discussion}\label{sec:discussion}

In view of the author, the major result of this paper is the realization that
the atomic scale stress correlation function, discussed previously in Ref.\cite{Levashov20111,Levashov2013,Levashov2014B,Levashov20141},
contains in itself not only the vibration contribution to viscosity, 
as it has been assumed, but also the contribution from the structural relaxation.

We note that in some recent papers importance of the propagating stress waves for the glass
transition has been discussed \cite{Trachenko20091,Trachenko20161}.
Correlations between the local rearrangement events and viscosity also
have been addessed \cite{Iwashita20131,Ashwin20151}.

The interpretations suggested here appears to be well aligned with
the philosophy behind the mode-coupling theory (MCT) where it is assumed that the memory functions 
related to the evolutions of the correlations functions can be split into the rapidly decaying binary collision part and a slowly decaying mode-coupling
part that couples the decay of the considered correlation functions to the
density fluctuations, i.e., to the static structure factor and to the intermediate scattering function(s) \cite{Gotze19841,Leutheusser19841,Kirkpatrick19841,Kirkpatrick19861,Marchetti19921,Balucani19871,Balucani19881,Gotze1998sw,Balucani19901,Marchetti19921,Balucani19931,Balucani1994book,Balucani2001,Das20021,Egorov20081,Bryk20161,Gonzales20171}.With respect to the transverse current correlation function and the macroscopic Green-Kubo stress correlation function, this approach has been described in Ref.
\cite{Kirkpatrick19841,Kirkpatrick19861,Marchetti19921,Balucani19871,Balucani19881,Gotze1998sw,Balucani19901,Marchetti19921,Balucani19931,Balucani1994book,Balucani2001,Das20021,Egorov20081,Bryk20161,Gonzales20171}. 
In particular, it allows obtaining a very good fit of the simulated macroscopic
stress correlation functions for alkali metals near the melting point with the
curves calculated within the MCT approach \cite{Balucani19931}. 
At present, various issues related to this approach are still discussed in the literature \cite{Balucani2001,Das20021,Egorov20081,Bryk20161,Trachenko20161,Gonzales20171}. 
 
In our view, the data presented here might offer some additional insight into
the nature of the mode-coupling memory 
term discussed in the MCT approach.
In particular, the data presented in Fig.\ref{fig:FqtA-01},\ref{fig:ss-01-macro} suggest that the binary collision term should not be of importance for the times larger than $t \gtrapprox 0.5\tau$. 
Further, the results suggest that for 
the times $t \gtrapprox 0.5\tau$ the stress correlation functions are still formed by two distinct contributions.
One contribution can be associated with the propagating longitudinal and shear waves while another contribution is associated with the structural relaxation \cite{Gotze1998sw}.
From this perspective, it is of interest to develop a better
understanding of how these features are related to the features of the mode-coupling memory term or to the van Hove correlation function. 
Ideally, it is desirable to have the MCT predictions for the behaviors of the atomic scale stress correlation functions. 

From the perspective of the potential energy landscape (PEL) approach, our results suggest the presence of non-trivial size effects in the stress correlation functions calculated on the inherent structures. This issue requires further investigations.

\section{Conclusion}\label{sec:conclusion}

We investigated the behavior of a binary system of repulsive particles from 
the perspective of the atomic scale stress correlation functions that are directly 
relevant to the Green-Kubo expression for viscosity.
In this paper, we studied the behavior of the stress correlation functions at the temperatures which are deeper in the supercooled liquid range than the temperatures that we studied previously for the single component systems.

The major result of this paper is the elucidation of the contribution from the structural relaxation to the atomic scale stress correlation functions relevant for the Green-Kubo expression for viscosity.
Analysis of the results shows that the contribution from the structural relaxation to viscosity can be accounted through the consideration of the atomic stress autocorrelation function 
(there is also a relatively small contribution associated with the nearest neighbor shell).
However, this does not mean, in our view, that viscosity is a local quantity from the perspective of the structural relaxation.
The situation that it is sufficient to consider only the contributions from the stress autocorrelation function and the nearest neighbor shell
arises because positive and negative contributions from the larger distances mutually cancel each other. 

Besides the contribution to viscosity associated with the structural relaxation there is also the contribution associated with the shear waves \cite{Levashov20111,Levashov2013,Levashov20141}. 
The results presented here demonstrate the generality of the results obtained previously for a single component system \cite{Levashov20111,Levashov2013,Levashov20141}. 
In particular, we again observed the shear waves propagating over the large distances and demonstrated their relation to viscosity.
We found that the contribution from the shear waves 
can be observed for the times which are much larger
than it takes for the stress waves to propagate through the system. 
The results also show the important role of the periodic boundary conditions 
for understanding the structure of the stress correlation functions.

In comparison to the previous publications, we also studied the propagation of 
the atomic scale pressure-pressure correlation waves.
It was found that the pressure-pressure correlation waves are much more pronounced 
than the shear-shear correlation waves.
The obtained results also show that the attenuation of the pressure waves is 
less significant than the attenuation of the shear waves.
Besides, the results suggest that there might exist a structural pressure-pressure correlation range. This issue requires, in our view, further investigations.

Analysis of the stress correlation functions calculated on the inherent structures of different sizes suggests the presence on non-trivial size effects in the considered correlation functions.
This issue also requires further considerations.

\section{Acknowledgments}\label{sec:acknowledgements}

We are grateful to the second referee of this paper for bringing to our attention several highly relevant papers on the Mode-Coupling Theory.

We are grateful to the Supercomputing Center of the Novosibirsk State University for provided computational resources.


\end{document}